\documentclass[twoside]{ilcws08}
\usepackage[latin1]{inputenc}
\usepackage[dvips]{graphicx,epsfig,color}
\usepackage{wrapfig,rotating}
\usepackage{amssymb,amsmath,array}

\begin{document}
\title{Analysis of Tau-pair process in the ILD reference detector model}
\author{\Large{Taikan Suehara}
\vspace{.3cm}\\
The University of Tokyo - International Center for Elementary Particle Physics (ICEPP) \\
7-3-1 Hongo, Bunkyo-ku, Tokyo, 113-0033, Japan
}

\maketitle

\begin{abstract}
Tau-pair process has been analyzed in the ILD detector model as a benchmark process for LoI.
Results of background rejection, forward-backward asymmetry and polarization measurements
are obtained with full detector simulation.
\end{abstract}

\section{Goals for LoI}

Tau-pair process (e$^+$e$^-\rightarrow$Z$^\ast,\gamma\rightarrow\tau^+\tau^-$) at $\sqrt{\mathrm{s}} = 500$ GeV
is one of the benchmark processes\cite{benchmark} proposed by Research Director.
According to the report, this process is a good sample to examine detector performances of
\begin{itemize}
	\item tau reconstruction, aspects of particle flow,
	\item $\pi_0$ reconstruction,
	\item tracking of very close-by tracks.
\end{itemize}
In this process, tau leptons are highly boosted ($\gamma\sim140$), thus decay daughters
(mainly charged and neutral pions, muons and electrons) are concentrated in a very narrow angle. 
Reconstruction of $\pi_0$ from two photons is especially challenging for the ILC detectors.

Observables for the LoI are cross section, forward-backward asymmetry and polarization of tau leptons.
The polarization measurement requires identification of tau decays, including reconstruction of $\pi_0$.
Efficiency and purity of event selection cuts should be also a good measure of detector performance.

For physics motivation, tau-pair process is important as a precision measurement of the electroweak theory.
For example, measuring cross section and forward-backward asymmetry of tau-pair process very precisely
can probe existence of heavy Z' boson.

\section{Analysis framework and events}

\subsection{Event samples}
\label{sec:event}

Events of ILD\_00 LoI mass production\cite{masspro} are used for this study.
Events reconstructed and listed at DESY by approximately end of February are used in this analysis.
10.3 M SM events generated in SLAC are processed for background estimation with appropriate event weight.

Since the SLAC events have a polarization issues for tau-pair events,
tau-pair events generated in DESY are used instead of SLAC events in this analysis.
For other modes including tau, SLAC events are used.
Whizard 1.51 and TAUOLA\cite{tauola} are used to generate the DESY events.
Statistics of the signal channel is 500 fb$^{-1}$ both for e$^-_\mathrm{L}$e$^+_\mathrm{R}$
and e$^-_\mathrm{R}$e$^+_\mathrm{L}$ (total 2.3 M events).

Bhabha process (e$^+$e$^-$ elastic scattering) is an important background for tau-pair analysis.
Since the cross section of Bhabha process is too large ($\sim$ 17 nb for each polarization in SLAC events),
following preselection is applied to the SLAC events before simulation.
\begin{itemize}
	\item $|\cos\theta|$ of electron or positron must be smaller than 0.96.
	\item Opening angle between electron and positron must be larger than 165 deg.
\end{itemize}
After the preselection, the cross section is reduced to 50-90 pb.
$\sim$1.0 fb$^{-1}$ of preselected Bhabha events are simulated.

Preselection is also applied to $\gamma\gamma \to \tau\tau$ events with following cuts:
\begin{itemize}
	\item Opening angle between two taus must be larger than 170 deg.
	\item Energy sum of two taus is greater than 30 GeV.
\end{itemize}
The total cross section after the cuts is around 18 pb.
Around 100 k events passing preselection are processed.

Integrated luminosity is assumed to be 500 fb$^{-1}$ each
for two polarization setups, e$^-_\mathrm{L}$e$^+_\mathrm{R}$
and e$^-_\mathrm{R}$e$^+_\mathrm{L}$.
Assumed polarization ratio is 80\% for electron and 30\% for positron
(i.e.~for e$^-_\mathrm{L}$e$^+_\mathrm{R}$ setup 90\% of electrons are
leftly polarized and 65\% of positrons are rightly polarized).

\subsection{Tau clustering}

For tau clustering, an original clustering processor (TaJet) is applied to
the output of PandoraPFA. Following is a procedure of the processor.
\begin{enumerate}
	\item Sort particles in energy order.
	\item Select the most energetic charged particle (a tau candidate).
	\item Search particles to be associated to the tau candidate. Criteria is:
	\begin{enumerate}
		\item Opening angle to the tau candidate is smaller than 50 mrad., or
		\item Opening angle to the tau candidate is not larger than 1 rad.~and invariant mass with the tau candidate is less than 2 GeV (m$_\tau$ = 1.777 GeV). 
	\end{enumerate}
	\item Combine energy and momentum of the tau candidate and associated particle and treat the combined particle as the new tau candidate.
	\item Repeat from 3.
	\item After all remaining particles do not meet the criteria, remaining most energetic charged particle is the next tau candidate. (Repeat from 2.)
	\item After all charged particles are associated to tau candidates, remaining neutral particles are independently included in the cluster list as neutral fragments.
\end{enumerate}

In the clustering stage, events with $>$ 6 tracks are pre-cut to accelerate clustering since $>$ 99\% of tau decays have $\leq$ 3 charged particles.
Event with only one positive and one negative tau clusters are processed with latter analysis.

\section{Background suppression}

Main background of tau-pair analysis is Bhabha (e$^+$e$^-\rightarrow$e$^+$e$^-$), WW $\rightarrow$ $\ell\nu\ell\nu$ and $\gamma\gamma\rightarrow\tau^+\tau^-$.
Since cross sections of Bhabha and two-photon events are huge (about $10^4$ and $10^3$ larger than signal, respectively),
we need tight selection cuts for those background events.
Following cuts are applied to signals and all SM background events after the tau clustering.

\begin{enumerate}
	\item Number of tracks $\leq$ 6. Included as a pre-cut in tau clustering processor.
	\item Only one positive and one negative tau clusters must exist in the event.
				(Neutral clusters are allowed.)
	\item Opening angle of two tau candidates must be $>$ 178 deg.

		This cut efficiently suppresses WW $\rightarrow$ $\ell\nu\ell\nu$ background.
	\item ee and $\mu\mu$ events are rejected.

		Charged particles depositing $>$ 90\% of their energy in ECAL are identified as electrons, and
		charged particles depositing $<$ 70\% of their energy (estimated by curvature of their tracks) in ECAL+HCAL are identified as muons.
		Events with two electrons or two muons are rejected in this cut.
		This cut is especially for suppressing Bhabha and e$^+$e$^-\rightarrow\mu^+\mu^-$ events. Signal loss is about 6\%.
	\item $|\cos\theta| < 0.95$ for both tau clusters.

		t-channel Bhabha events are almost completely suppressed by this cut. 20\% of signal events are lost.
	\item $70 < \mathrm{E}_\mathrm{vis} < 450$ GeV. $\mathrm{E}_\mathrm{vis}$ does not include energy of neutral clusters.

		Lower bound suppresses $\gamma\gamma\rightarrow\tau^+\tau^-$ events, and upper bound suppresses Bhabha events. Signal lost is negligibly small.

\end{enumerate}

\begin{table}
\begin{center}
\scriptsize{
\begin{tabular}{|r|r|r|r|r|r|r|r|}
\hline
Cuts & Tau-pair & Bhabha & $\mu\mu$ & n$\ell$ + n$\nu$ & $\gamma\gamma\rightarrow\ell\ell$ & other $\gamma\gamma$, e$\gamma$ & other \\ \hline\hline
\# tracks, \# clusters          & 573180 & 2.88e+07 & 590770 & 1.15e+06	& 5.58e+08 & 4.07e+06 & 1.21e+06\\\hline
Opening angle $>$ 178 deg.      & 152865 & 1.89e+07 & 157430 & 7938     & 6.93e+06 & 59454    & 2633 \\\hline
$|\cos\theta| < 0.95$           & 142371 & 1.39e+07 & 147571 & 5020     & 6.25e+06 & 57746    & 610 \\\hline
ee, $\mu\mu$ veto               & 130383 & 96482    &	1606   & 3225     & 616265   & 45645    &	141 \\\hline
70 $<$ E$_\mathrm{vis}$ $<$ 450 GeV & 125400 & 5071 &	635    & 2953     &	1641     & 0        & 32  \\\hline
\end{tabular}
(a) e$^-_\mathrm{L}$ (80\%) e$^+_\mathrm{R}$ (30\%)

\begin{tabular}{|r|r|r|r|r|r|r|r|}
\hline
Cuts & Tau-pair & Bhabha & $\mu\mu$ & n$\ell$ + n$\nu$ & $\gamma\gamma\rightarrow\ell\ell$ & other $\gamma\gamma$, e$\gamma$ & other \\ \hline\hline
\# tracks, \# clusters          & 446551 & 2.68e+07 & 460874 & 116198 & 5.58e+08 & 46898050 & 1194395 \\\hline
Opening angle $>$ 178 deg.      & 127070 & 1.73e+07 & 133628 & 519    & 6.93e+06 & 59920    & 2934\\\hline
$|\cos\theta| < 0.95$           & 118426 & 1.23e+07 & 125113 & 326    &	6.25e+06 & 58987    & 512\\\hline
ee, $\mu\mu$ veto               & 108778 & 88385    &	1027   & 200    &	616265   & 46196    &	107\\\hline
70 $<$ E$_\mathrm{vis}$ $<$ 450 GeV & 103197 & 4857 & 383    & 183    & 1641     & 0        & 16 \\\hline
\end{tabular}
(b) e$^-_\mathrm{R}$ (80\%) e$^+_\mathrm{L}$ (30\%)

\caption{Cut statistics for background suppression. Preselection (See Section \ref{sec:event} for details) is applied for Bhabha events before these cuts. 
Number of events are normalized to 500 fb$^{-1}$.
The same statistics is used for (a) and (b): only event weighting is different.}
\label{tab:smcuts}
}
\end{center}
\end{table}

\begin{figure}
	\begin{center}
        \begin{minipage}{0.4\textwidth}
                \begin{center}
                        \includegraphics[width=1\textwidth]{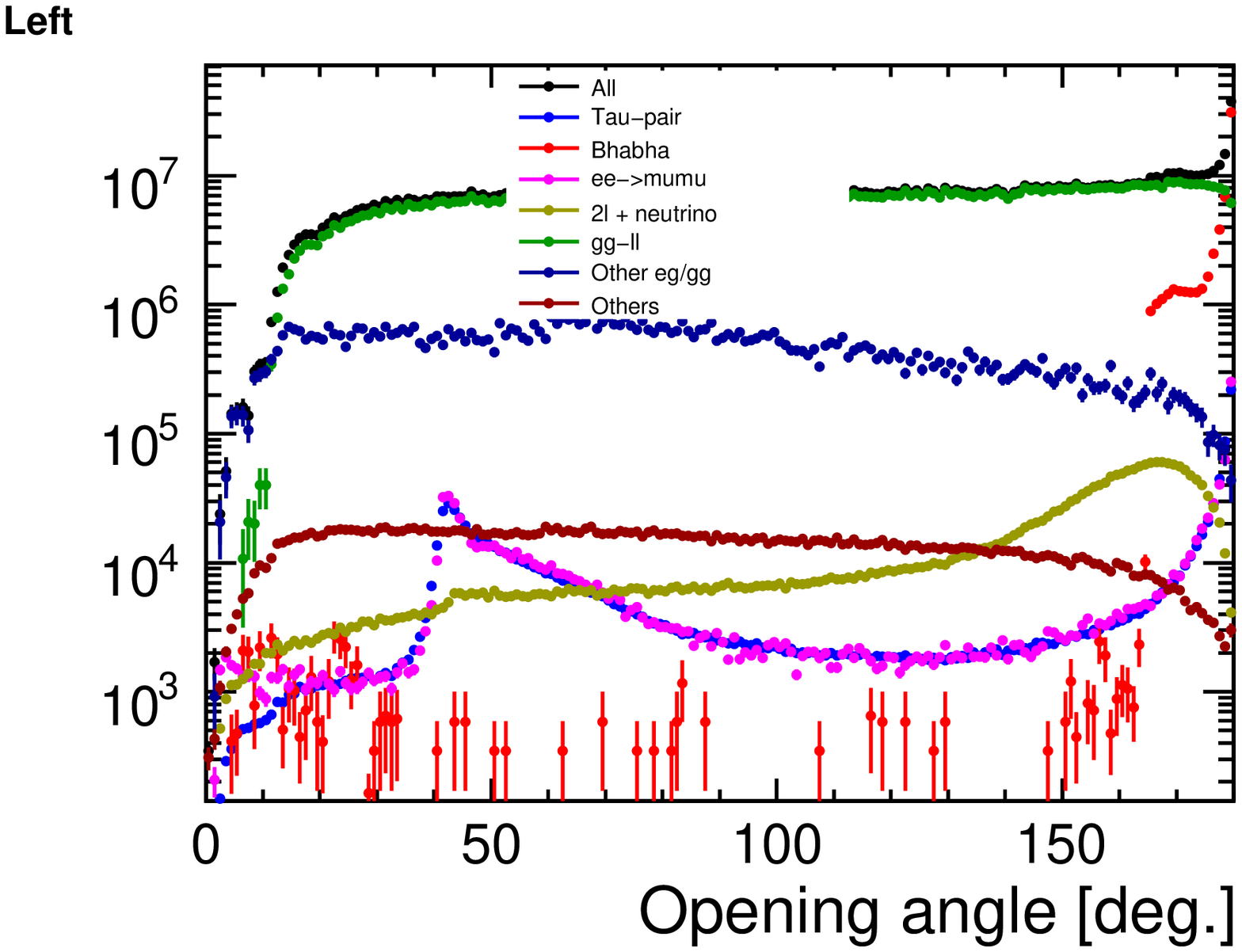}
                        \includegraphics[width=1\textwidth]{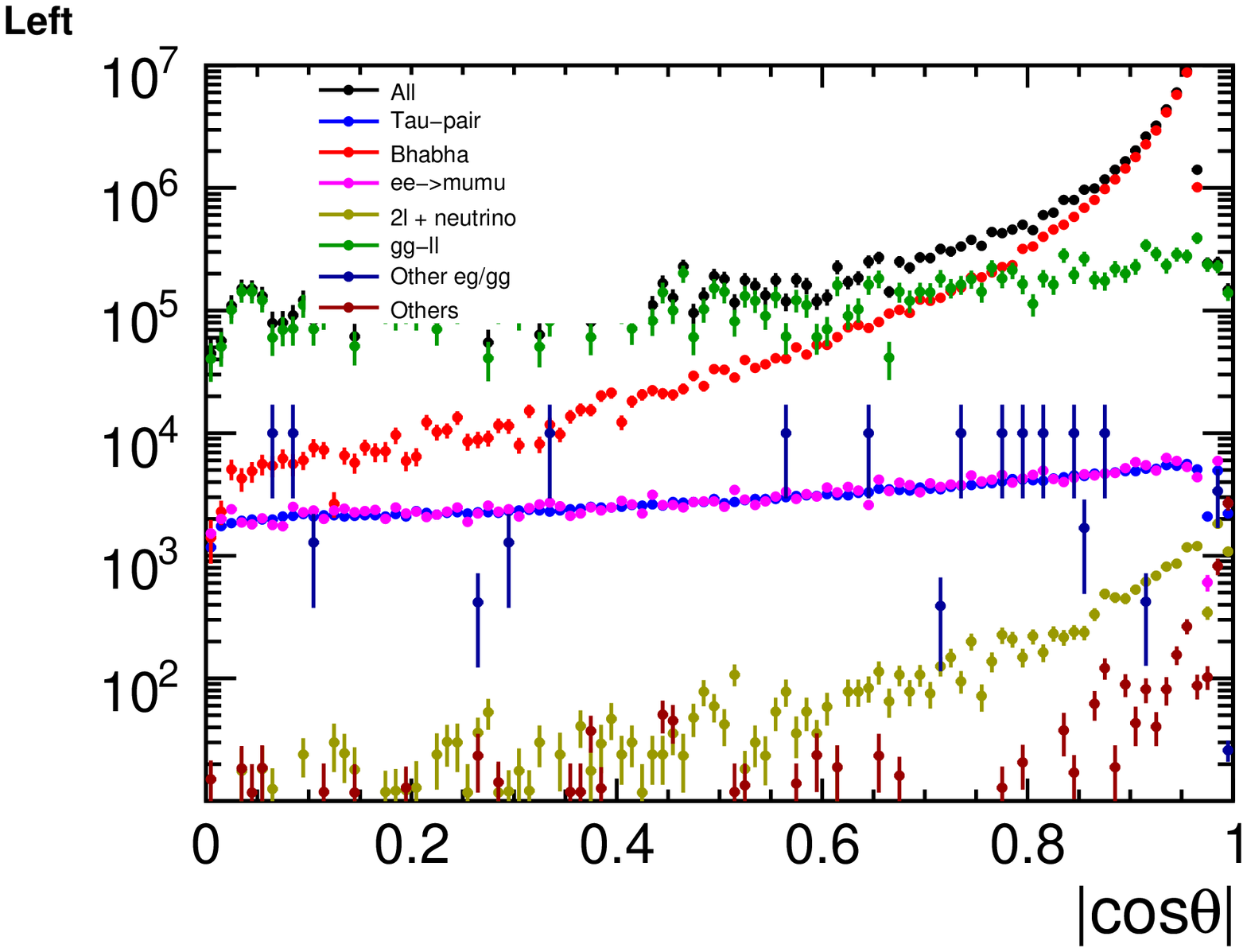}
                        \includegraphics[width=1\textwidth]{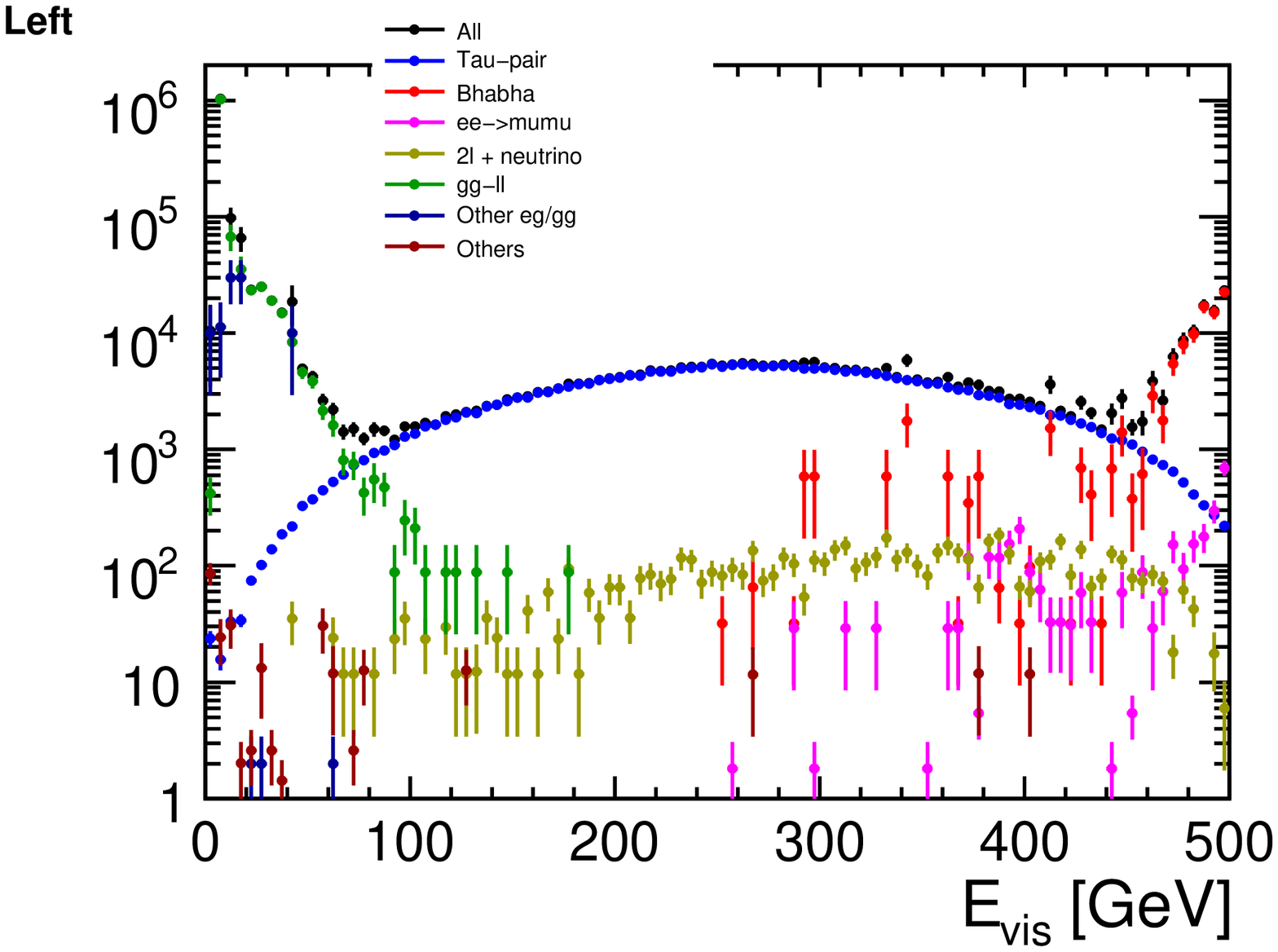}
                \end{center}
        \end{minipage}
        \begin{minipage}{0.4\textwidth}
                \begin{center}
                        \includegraphics[width=1\textwidth]{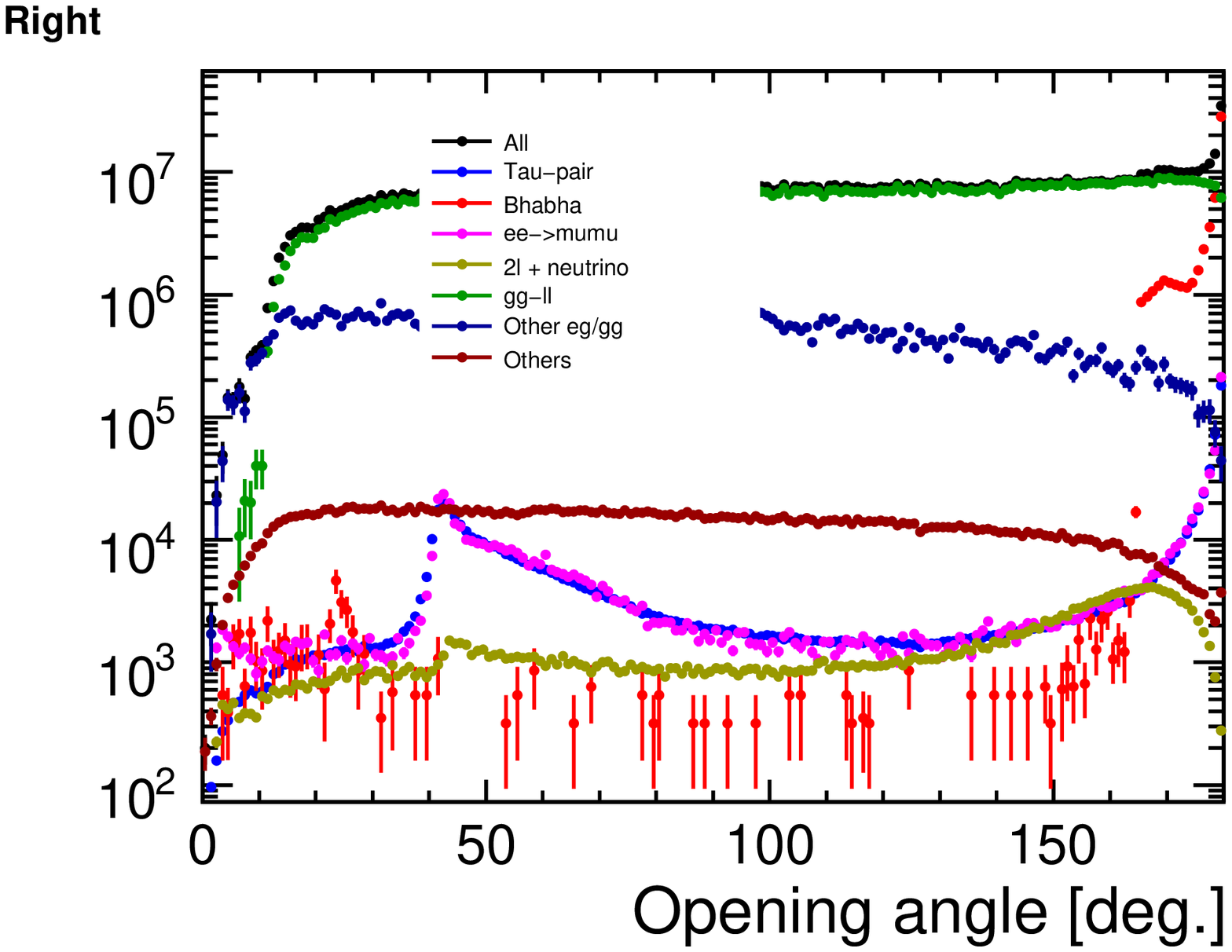}
                        \includegraphics[width=1\textwidth]{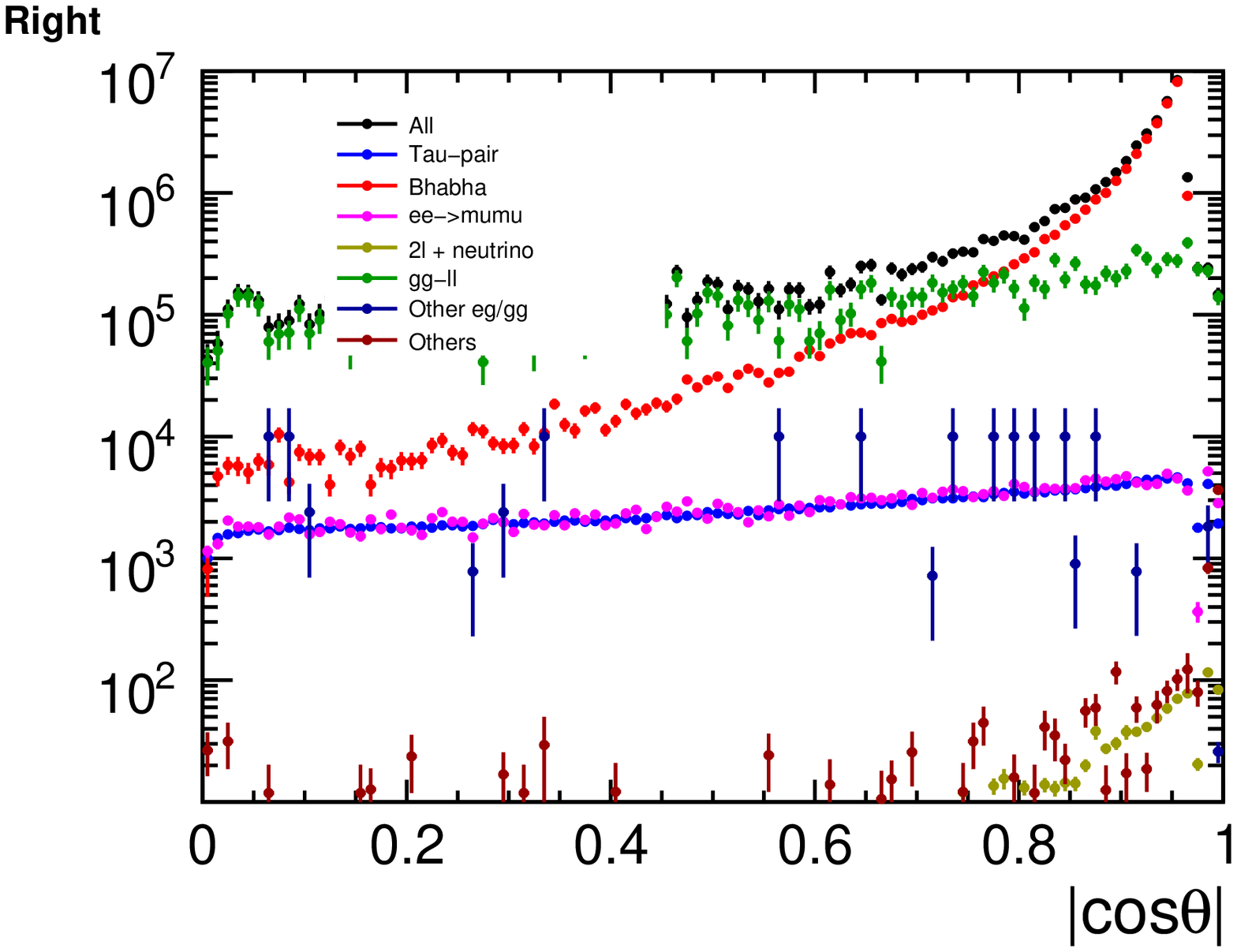}
                        \includegraphics[width=1\textwidth]{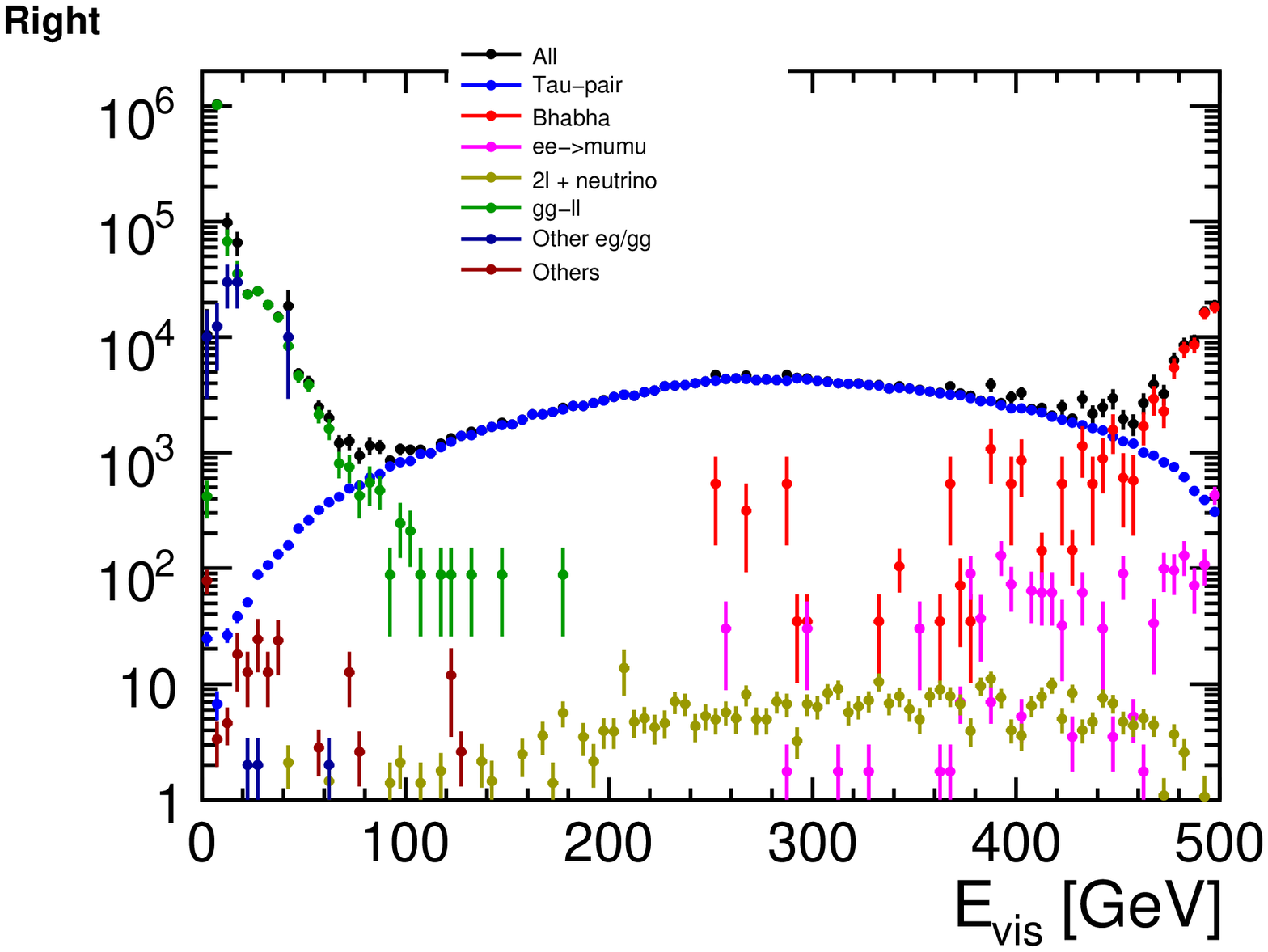}
                \end{center}
        \end{minipage}
        \caption{Distribution of cut values. Left column shows e$^-_\mathrm{L}$e$^+_\mathrm{R}$ distribution and
				Right column shows e$^-_\mathrm{R}$e$^+_\mathrm{L}$ distribution. Cuts are applied from top, and
				ee and $\mu\mu$ veto cuts are applied between second and third rows, whose distributions are omitted.}
        \label{fig:cuts}
	\end{center}
\end{figure}

Table \ref{tab:smcuts} shows the result of these cuts and Figure \ref{fig:cuts} shows distribution of cut values. 
Most of the background is effectively cut off by the cuts, $\sim$10\% level of the signal.
Remaining background is mainly Bhabha, $\gamma\gamma\rightarrow\tau\tau$ and WW$\rightarrow\ell\nu\ell\nu$.

Purity of tau selection is 92.4\% in e$^-_\mathrm{L}$e$^+_\mathrm{R}$ sample and
93.6\% in e$^-_\mathrm{R}$e$^+_\mathrm{L}$.
The difference is mainly from difference of the cross section between each polarization.

Selection efficiency of tau-pair events is literally low (15.8\% in e$^-_\mathrm{L}$e$^+_\mathrm{R}$ and
16.3\% in e$^-_\mathrm{R}$e$^+_\mathrm{L}$).
However, the `nocut' number contains radiative events, which have effectively lower $\sqrt{s}$ and should not be
used in the analysis. These radiative events are cut by the opening angle selection.
The real efficiency varies by the definition of the events.
The acceptance of `softly-radiated' tau-pair events is determined by the opening angle cuts.
Loosing the cut accepts more events, although $\gamma\gamma\rightarrow\tau\tau$
 and WW$\rightarrow\ell\nu\ell\nu$ background significantly increase.

\section{Cross section}

Cross section can be easily obtained by count-based method since background amount is low.
Assuming background subtraction can be performed in the error of statistics, we obtain
number of signal event as $125400 \pm 368$ (e$^-_\mathrm{L}$e$^+_\mathrm{R}$) and
$103197 \pm 332$ (e$^-_\mathrm{R}$e$^+_\mathrm{L}$), ie.~0.29\% and 0.32\% statistical error, respectively.
The statistical error is dominated by signal statistics, so poor statistics of background events 
in the current MC sample can only
have small effect on these numbers (0.30\% and 0.33\% statistical error, even if background is doubled).
Systematic error can be introduced by polarization error, MC incorrespondance to real detector etc., but
it cannot be accurately estimated in this stage of detector development and thus not considered now.

\section{Forward-backward asymmetry}

\begin{figure}
	\begin{minipage}[t]{0.47\textwidth}
	\begin{center}
		\includegraphics[width=0.95\columnwidth]{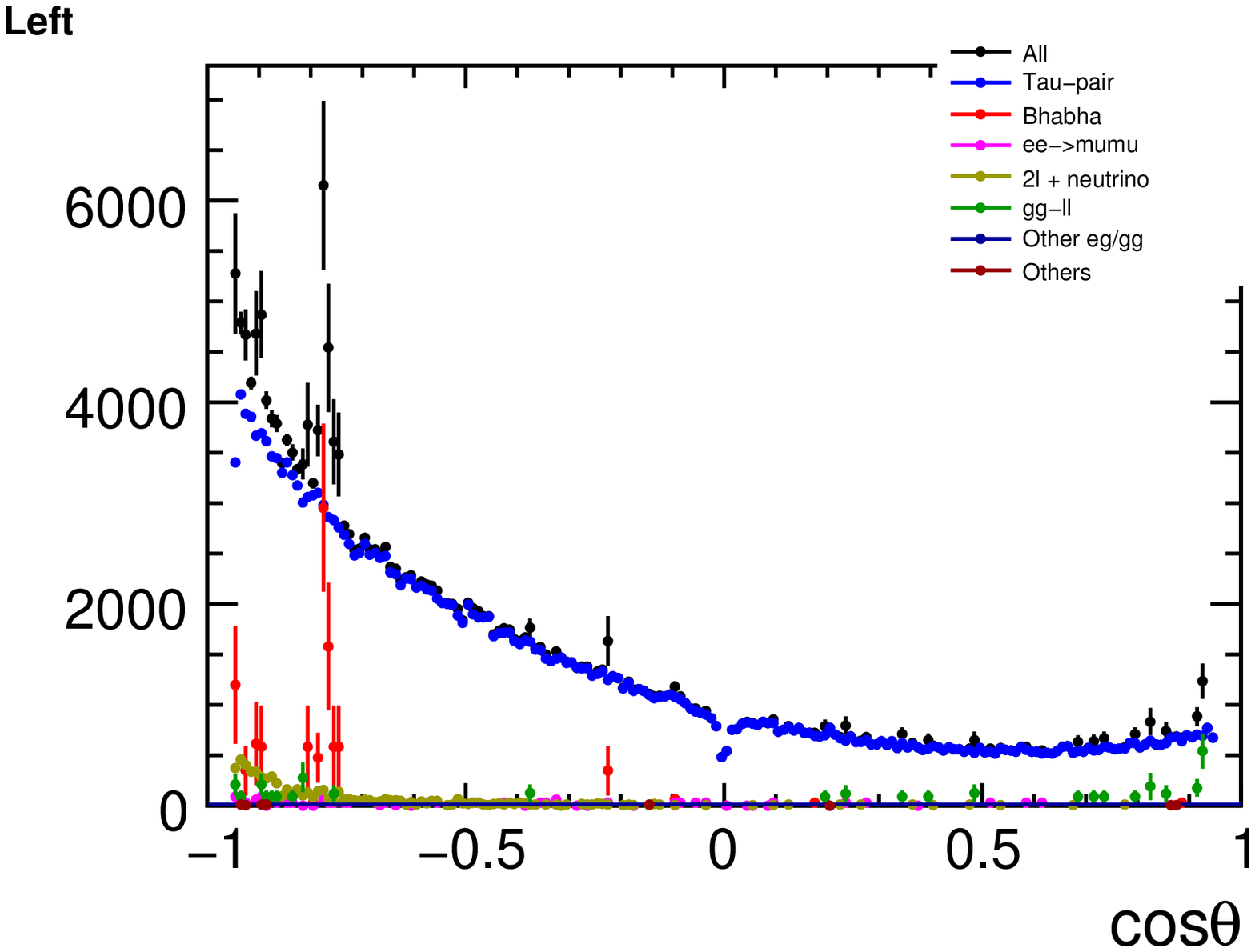}
		\small{(a) e$^-_\mathrm{L}$ (80\%) e$^+_\mathrm{R}$ (30\%)}
	\end{center}
	\end{minipage}
	\hfill
	\begin{minipage}[t]{0.47\textwidth}
	\begin{center}
		\includegraphics[width=0.95\columnwidth]{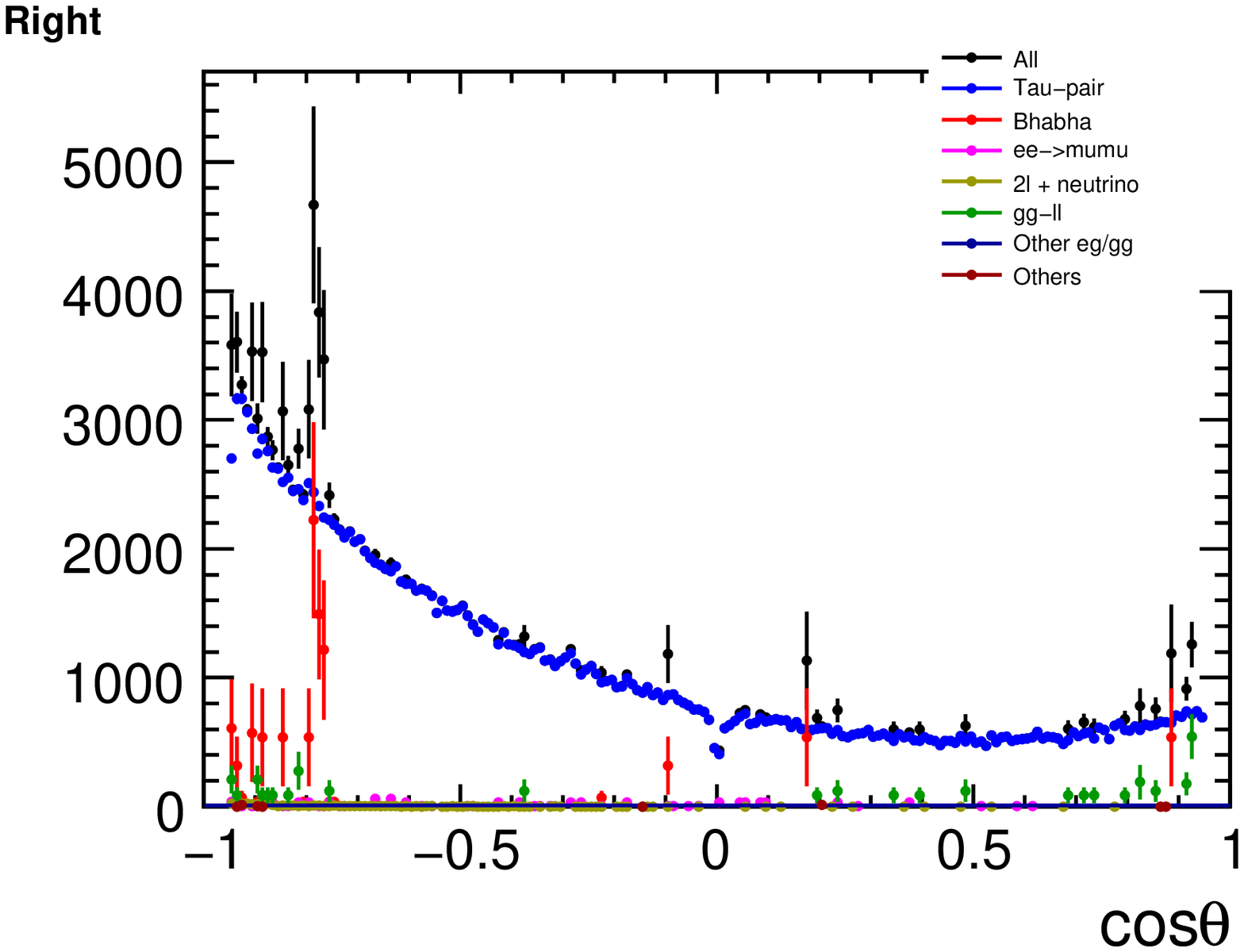}
		\small{(b) e$^-_\mathrm{R}$ (80\%) e$^+_\mathrm{L}$ (30\%)}
	\end{center}
	\end{minipage}
\caption{Angular distribution of $\tau^+$ momentum direction.
Number of events are normalized to 500 fb$^{-1}$.
Error bars stand for statistical errors in current MC statistics.
The same statistics is used for (a) and (b): only event weighting is different.
}
\label{fig:afb}
\end{figure}

Figure \ref{fig:afb} shows a result on angular distribution of $\tau^+$ leptons
($\tau^-$ events are essentially the same event-by-event since we require opening angle $>$ 178 deg.).

Assuming that background can be subtracted effectively, forward-backward asymmetry is calculated by following formulae.
\begin{eqnarray}
	A_{FB} &=& \frac{N_F - N_B}{N_F + N_B}, \\
	\sigma{}A_{FB} &=& \sqrt{\left(\frac{\partial{}A_{FB}}{\partial{}N_F}\sigma{}N_F\right)^2 + \left(\frac{\partial{}A_{FB}}{\partial{}N_B}\sigma{}N_B\right)^2}, \\
	\sigma{}N_F &=& \sqrt{N_F + N_{FBG}}, \quad \sigma{}N_B = \sqrt{N_B + N_{BBG}},
\end{eqnarray}
where $N_B$ is number of signal events in backward region ($\cos\theta<0$), $N_F$ is number of signal events in forward region ($\cos\theta>0$),
$N_{BBG}$ is number of background events in the backward region and $N_{FBG}$ is number of background events in the forward region.
The formulae can be reduced to
\begin{equation}
	\sigma{}A_{FB} = \frac{2\sqrt{N_B^2(N_F+N_{FBG}) + N_F^2(N_B+N_{BBG})}}{(N_F+N_B)^2}.
\end{equation}

Result of the calculation is:
\begin{eqnarray}
	\mathrm{e}^-_\mathrm{L}\mathrm{e}^+_\mathrm{R}&:& N_F = 95529, N_B = 29872, N_{FBG} = 9201, N_{BBG} = 1130, A_{FB} = 52.36 \pm 0.25\%, \\
	\mathrm{e}^-_\mathrm{R}\mathrm{e}^+_\mathrm{L}&:& N_F = 75556, N_B = 27640, N_{FBG} = 5477, N_{BBG} = 1605, A_{FB} = 44.19 \pm 0.28\%.
\end{eqnarray}
Statistical accuracy of $A_{FB}$ is 0.48\% and 0.63\%, respectively.

\section{Decay mode separation}

Separating decay modes of tau is essential for the polarization measurement.
There are five dominant decay modes of tau, $\tau^+ \rightarrow \mathrm{e}^+\overline{\nu_\mathrm{e}}\nu_\tau$ (17.9\%),
$\tau^+ \rightarrow \mu^+\overline{\nu_\mu}\nu_\tau$ (17.4\%),
$\tau^+ \rightarrow \pi^+\nu_\tau$ (10.9\%),
$\tau^+ \rightarrow \rho^+\nu_\tau \rightarrow \pi^+\pi^0\nu_\tau$ (25.2\%),
and $\tau^+ \rightarrow \mathrm{a}_1^+\nu_\tau \rightarrow \pi\pi\pi\nu_\tau$ (9.3\% (1-prong) and 9.0\% (3-prong)).
Other decay modes (10.3\%) include Kaons and multi-pions in other resonant modes or continuum.

We utilize a neural network for the decay mode selection.
Two separate networks are trained for 1-prong and 3-prong events.
1-prong decay includes $\mathrm{e}^+\overline{\nu_\mathrm{e}}\nu_\tau$, $\mu^+\overline{\nu_\mu}\nu_\tau$,
$\pi^+\nu_\tau$, $\rho^+\nu_\tau$ and $\mathrm{a}_1^+\nu_\tau$ modes.
Input variables of the 1-prong neural net are as follows.
\begin{itemize}
	\item Two lepton-ID values. Likelihood-based lepton ID software was developed, but due to the known issues of
the event production the lepton ID is not properly worked on the mass production sample.
As a simpler lepton ID, we use ratio between the energy deposit of the electromagnetic calorimeter (ECAL)
and the total deposit energy for the electron ID,
and ratio between the calorimeter energy deposit and the track momentum for the muon ID. These two variables are included in the neural network.
(Variable (a) and (b).)
	\item Energy of the charged particle and two kinds of energy sums of the neutral particles.
	The neutral energy sums contain particles whose ECAL energy deposit is $>$ 80\%, and $<$ 80\% of the total energy deposit, respectively.
	Particles with ECAL energy deposit $<$ 80\% are considered to be hadrons, which contain more spurious particles
	from a mis-fragmentation of energetic charged particles mainly at HCAL.
	The energy sums are especially expected to discriminate $\pi$ mode.
	(Variable (c), (d) and (e).)
	\item Number of neutral particles except neutral hadrons. Number of photons is a powerful information to
	separate $\rho$ (expected number of photons is 2) and a$_1$ (expected number of photons is 4).
	(Variable (f).)
	\item Invariant masses of all reconstructed particles except neutral hadrons and invariant masses of photons.
	Invariant masses of all reconstructed particles should equal to the masses of intermediate particles, $\rho$ and a$_1$.
	If photons are reconstructed properly, invariant masses of photons are close to that of $\pi_0$.
	For the photon / hadron separation, above criteria is used again.
	(Variable (g) and (h).)
	\item Energy of the third-energetic neutral particle. This variable is also to separate $\rho$ and a$_1$.
	Since $\rho$ can have at most two photons, energy of the third photon should be small even if it exists in the $\rho$ mode.
	(Variable (i).)
\end{itemize}

We use two hidden layers, first layer has 18 neurons and second has 9 neurons.
The output neurons are likelihood value of $\mathrm{e}^+\overline{\nu_\mathrm{e}}\nu_\tau$, $\mu^+\overline{\nu_\mu}\nu_\tau$,
$\pi^+\nu_\tau$, $\rho^+\nu_\tau$ and $\mathrm{a}_1^+\nu_\tau$ modes (5 neurons),
which is set to 1(true)/0(false) by the MC information in the training samples.

\begin{figure}
	\begin{center}
	\begin{minipage}{0.35\textwidth}
		\begin{center}
			\includegraphics[width=1\textwidth]{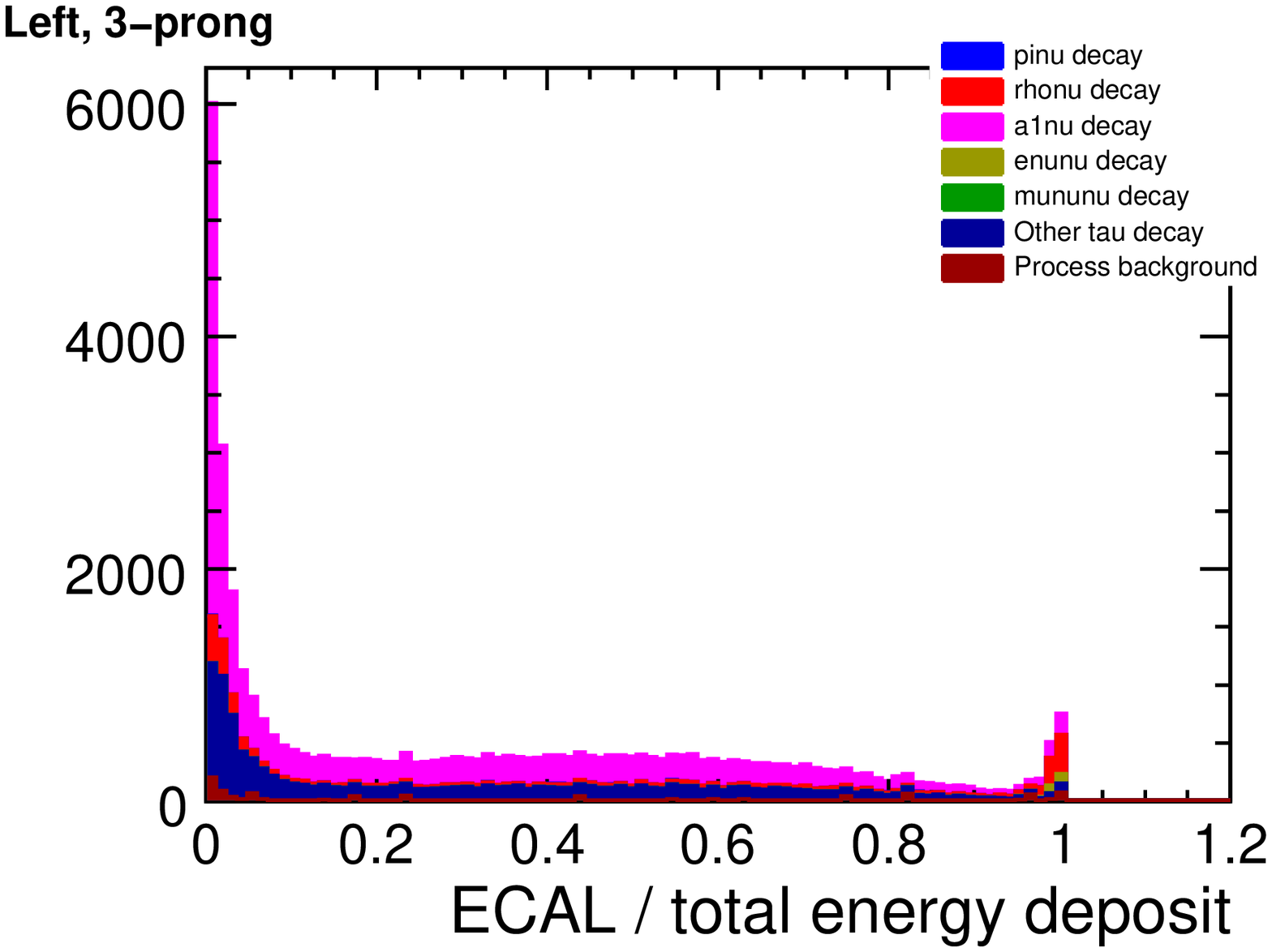}\par (a')
			\includegraphics[width=1\textwidth]{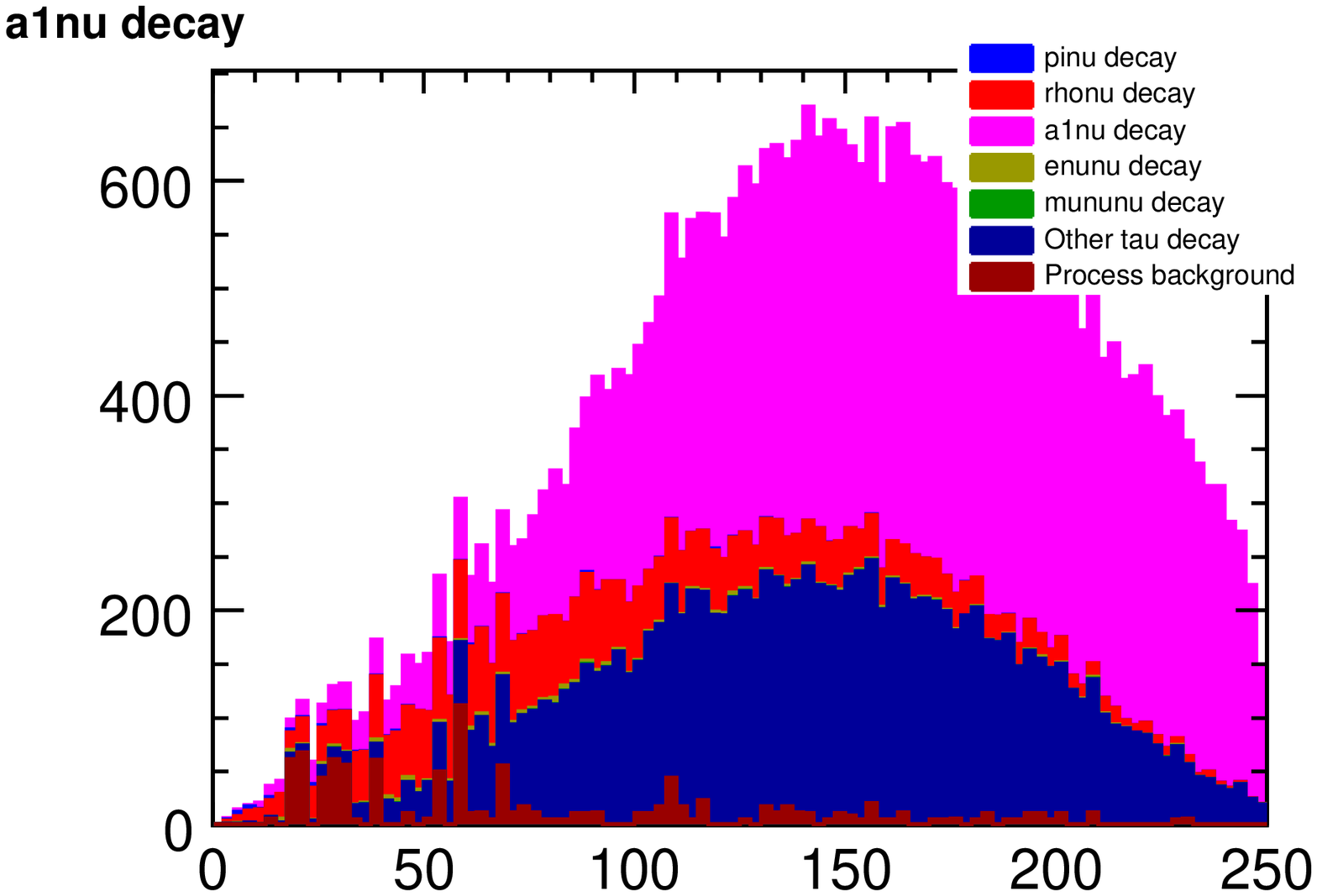}\par (c')
			\includegraphics[width=1\textwidth]{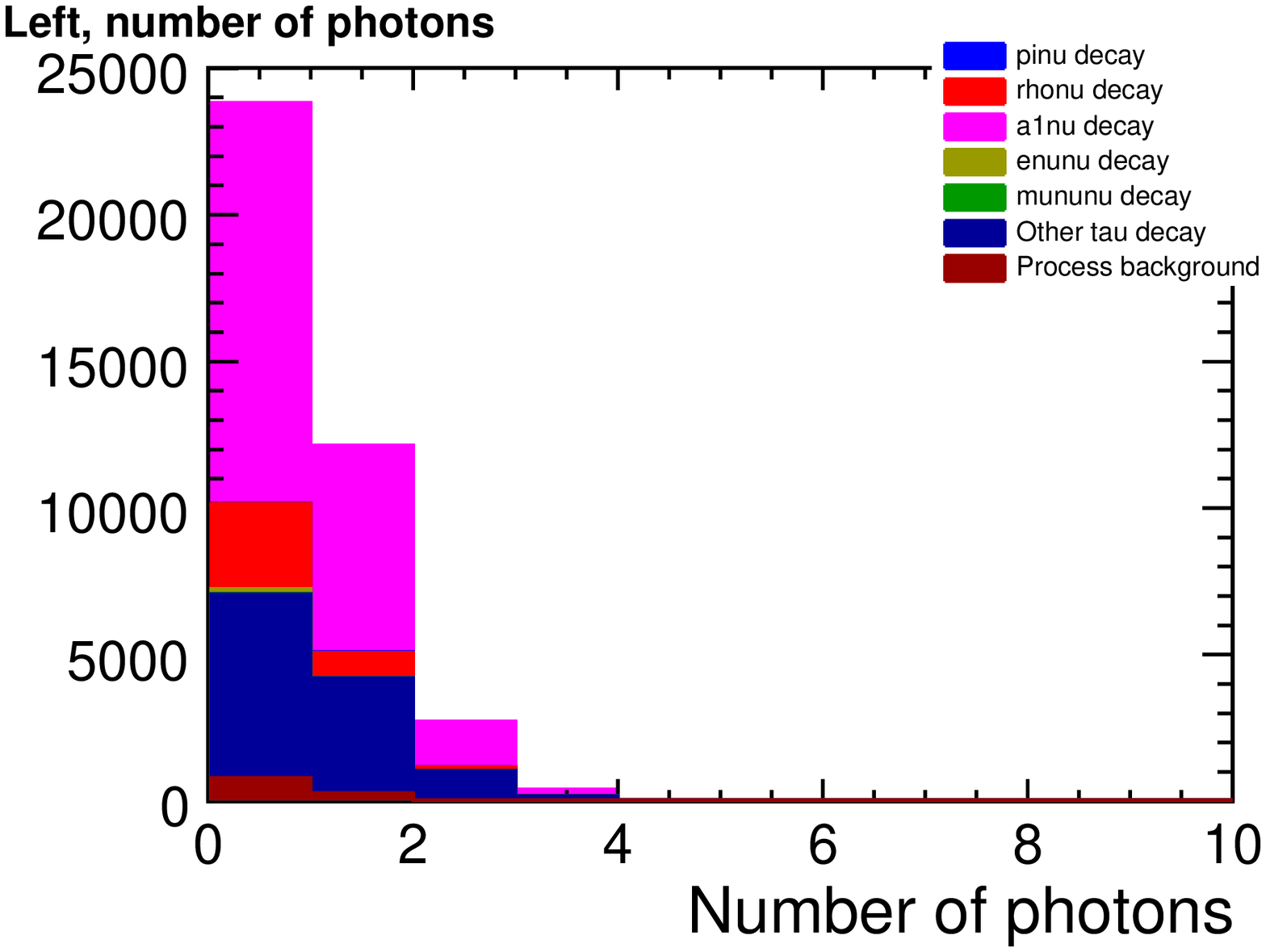}\par (f')
			\includegraphics[width=1\textwidth]{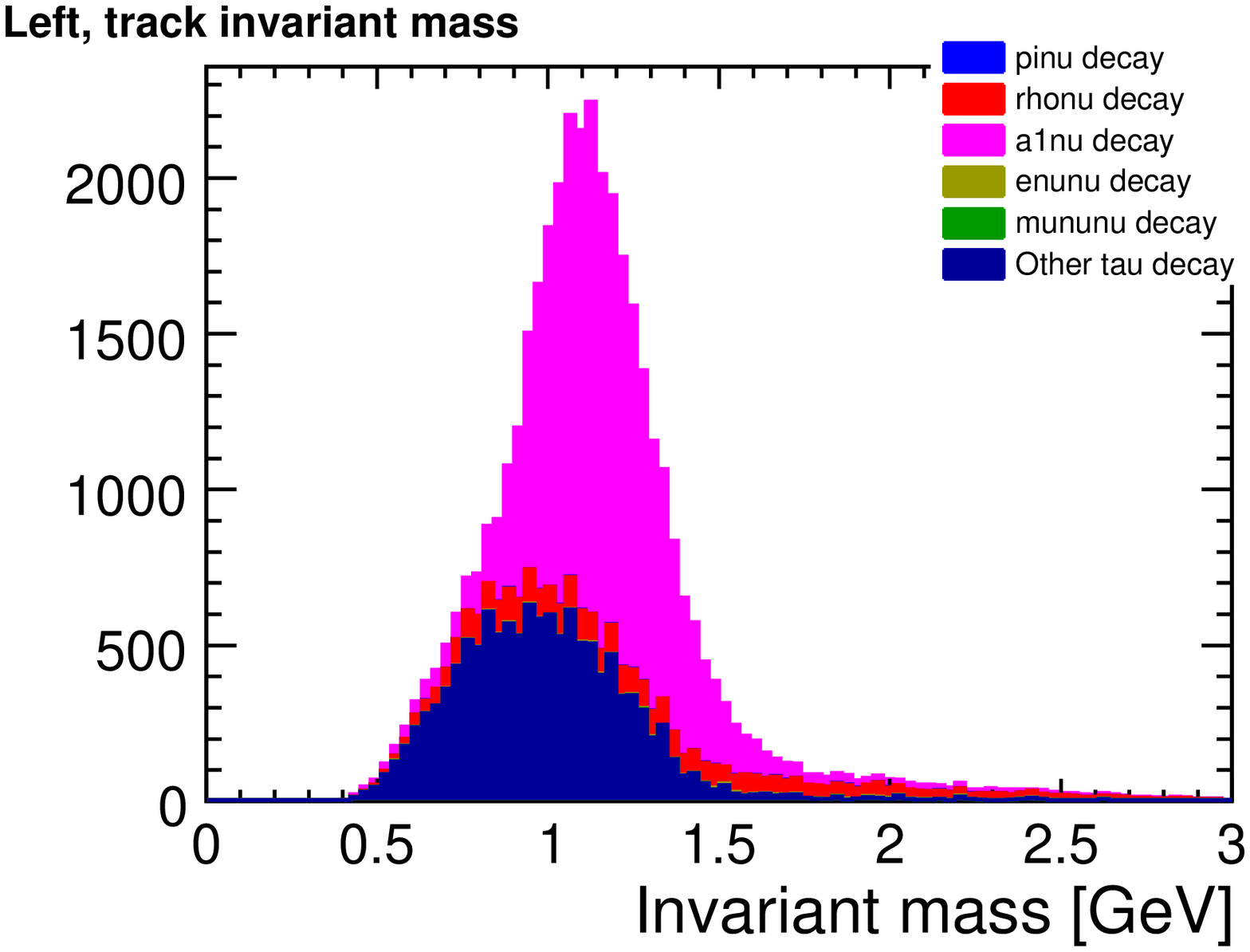}\par (j')
		\end{center}		
	\end{minipage}
	\begin{minipage}{0.35\textwidth}
		\begin{center}
			\includegraphics[width=1\textwidth]{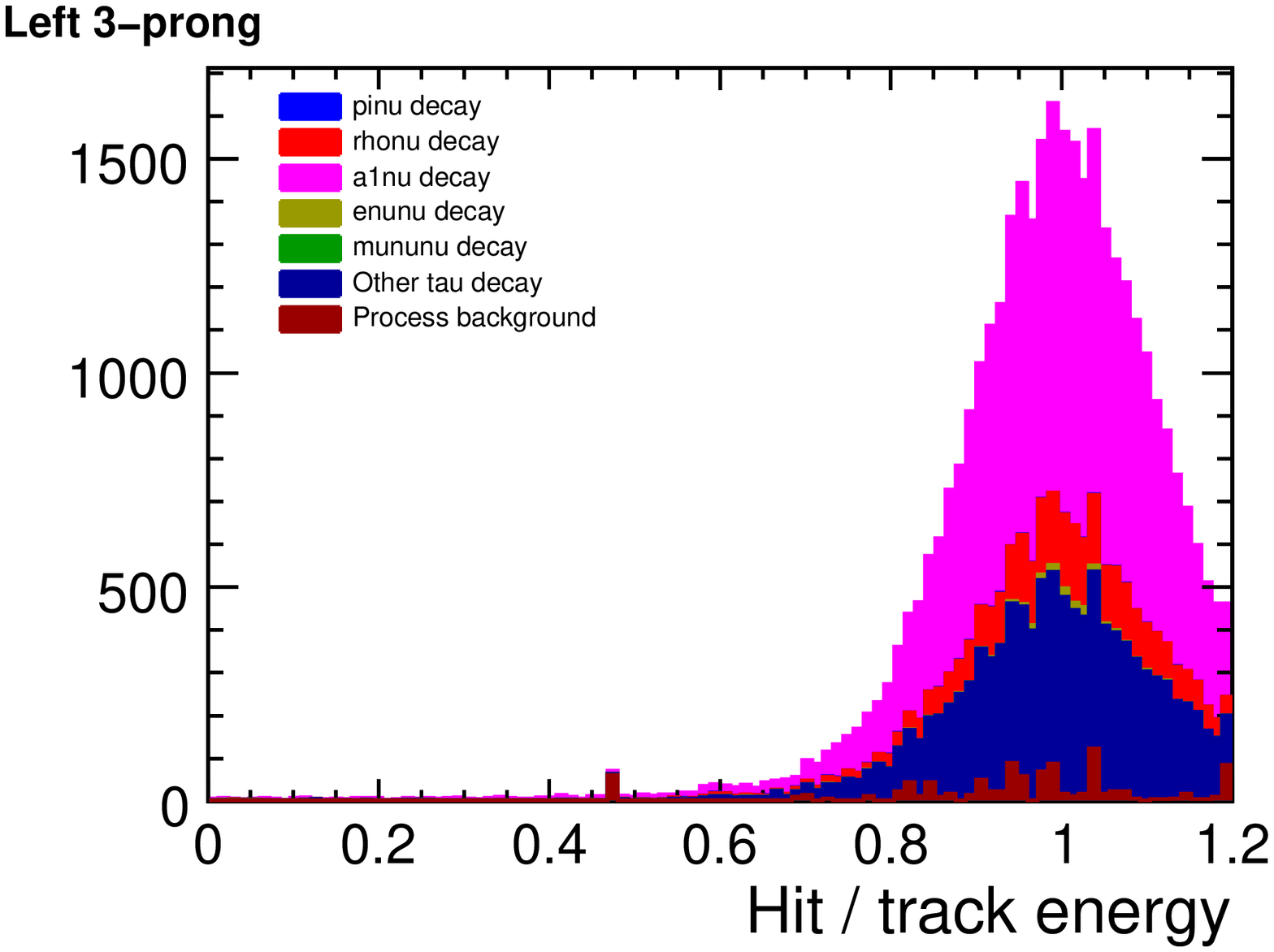}\par (b')
			\includegraphics[width=1\textwidth]{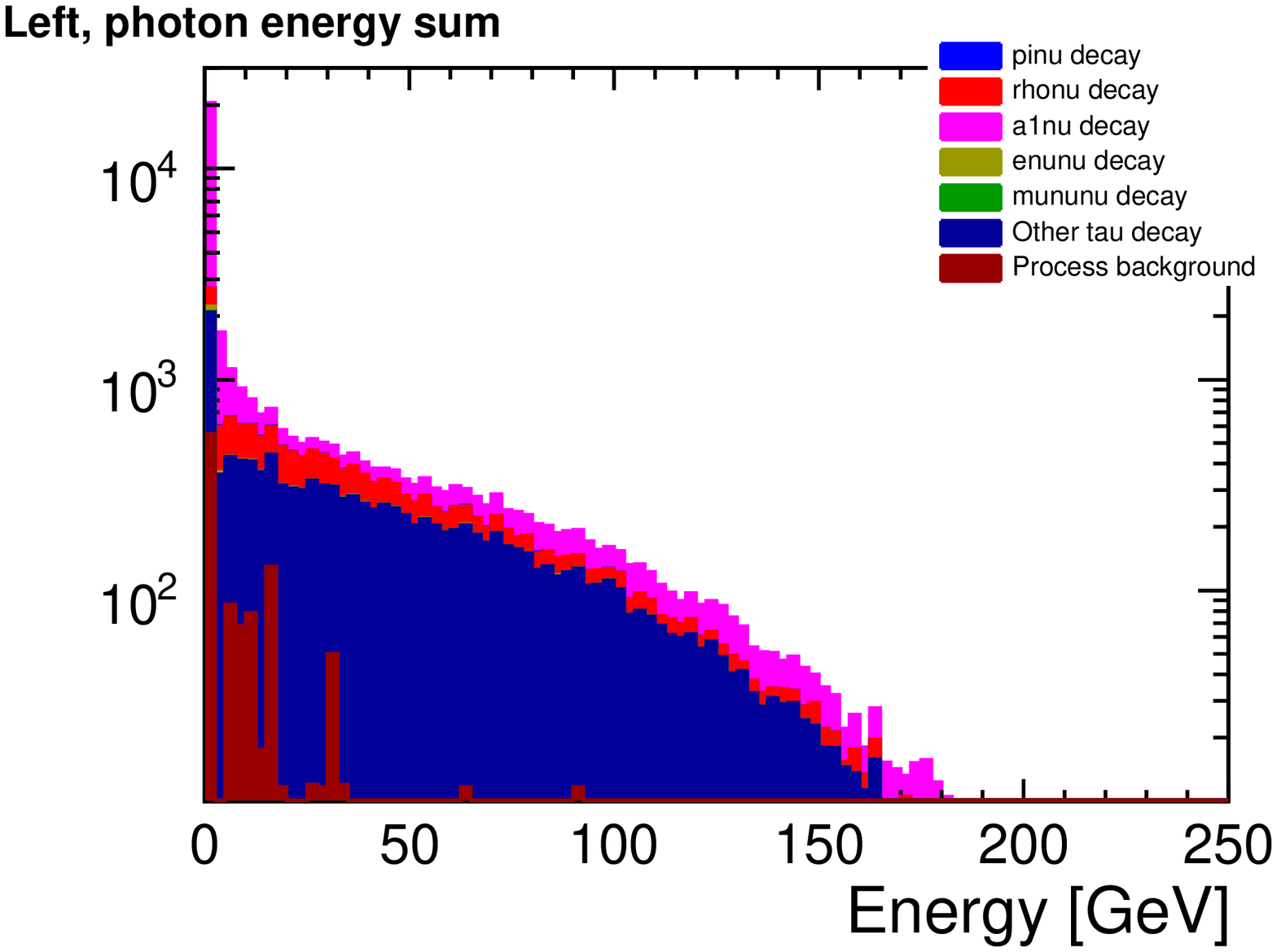}\par (d')
			\includegraphics[width=1\textwidth]{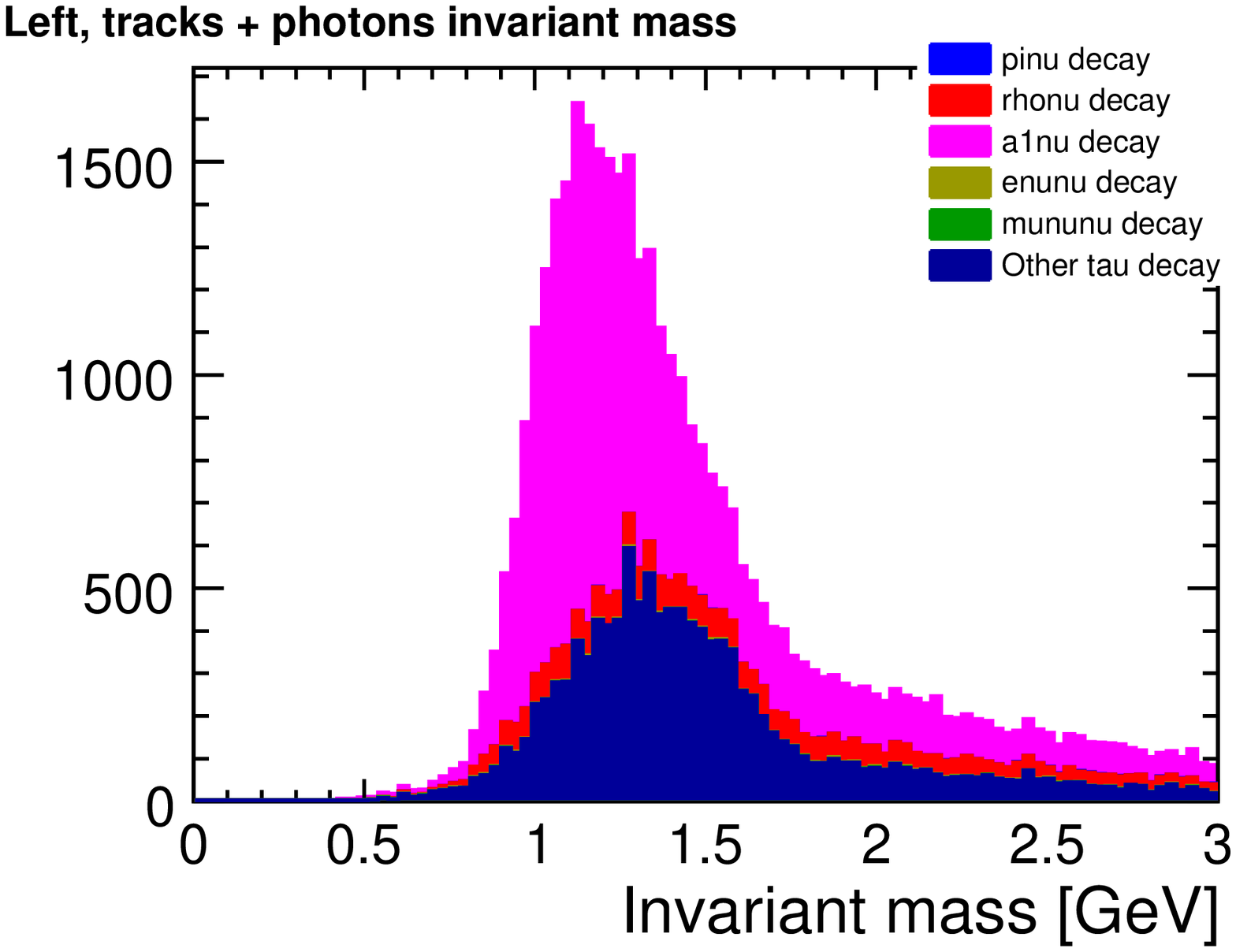}\par (g')
			\includegraphics[width=1\textwidth]{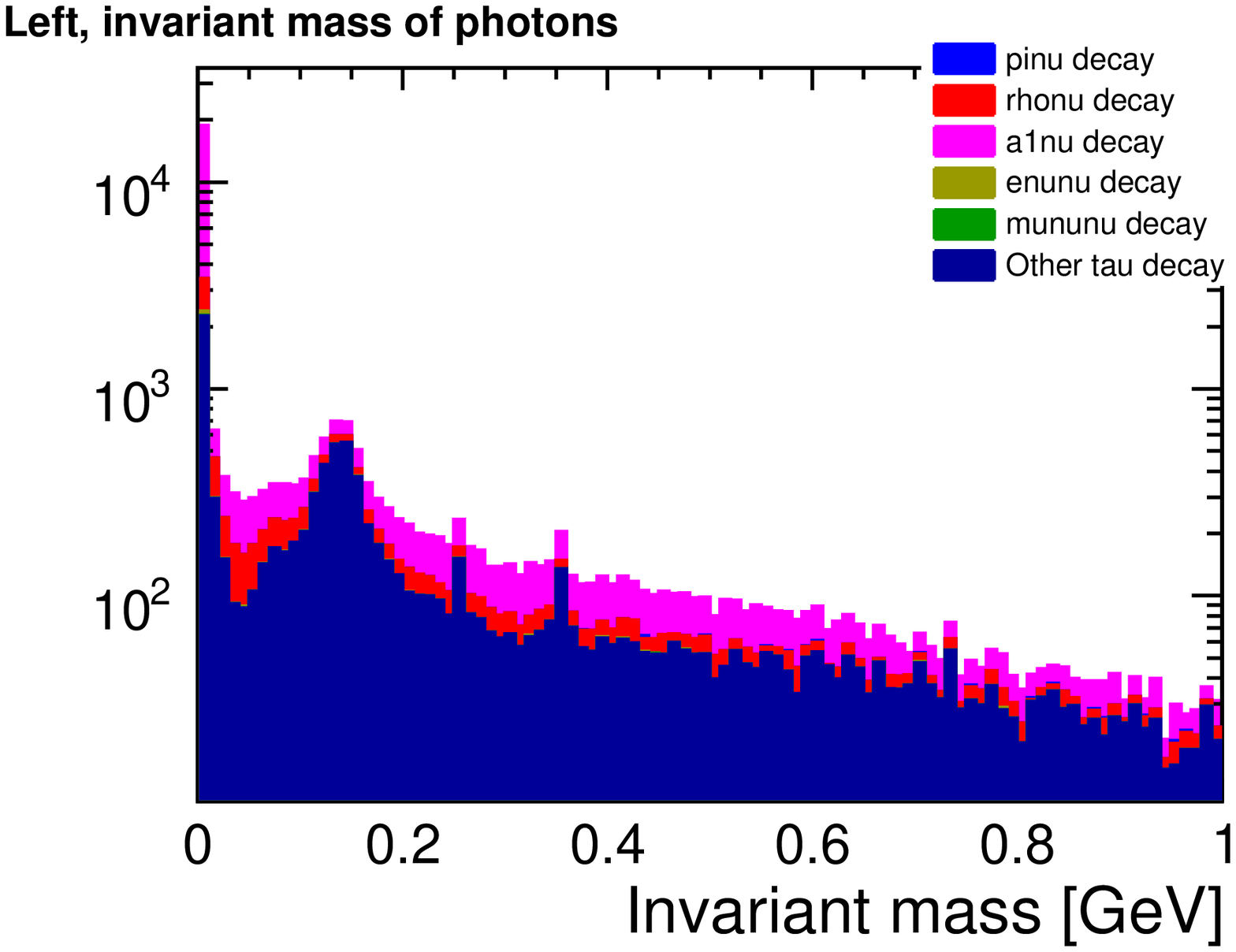}\par (h')
		\end{center}		
	\end{minipage}
	\end{center}		
	\caption{Distributions of the input variables for the 3-prong neural network.
e$^-_\mathrm{L}$ (80\%) e$^+_\mathrm{R}$ polarization is used for the plots.}
	\label{fig:inputvars3p}
\end{figure}

For the 3-prong events, only a$_1$ is the discriminating decay mode.
Input variables (a), (b), (c), (d), (f), (g), (h) (noted with (a')-(h') in Fig.~\ref{fig:inputvars3p})
in the 1-prong selection are also included in the 3-prong selection.
There is one additional variable, which is:
\begin{itemize}
	\item Invariant mass of all charged particles. This should equals to the mass of a$_1$ if the decay is a$_1$ mode.
	(Variable (j').)
\end{itemize}

We use two hidden layers with 16 and 5 neurons.
The only output neuron stands for likelihood value of a$_1$ mode, set to 1/0 in the training samples as well.

For the training, half of the tau-pair events in the mass production are used.
Number of epochs is 1000 for both 1-prong and 3-prong network.

\begin{figure}
	\begin{center}
	\begin{minipage}{0.35\textwidth}
		\begin{center}
			\includegraphics[width=1\textwidth,clip]{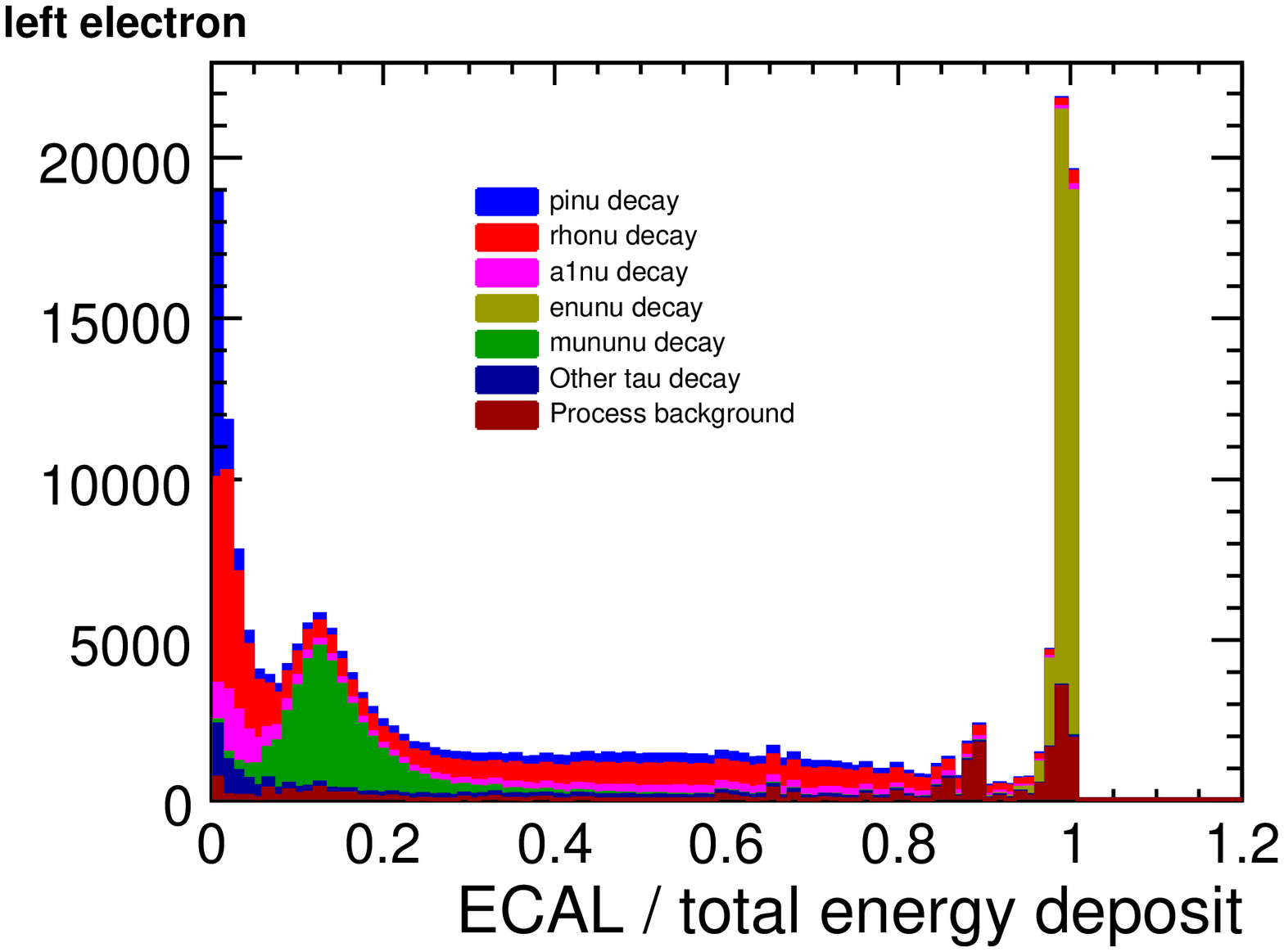}\par (a)
			\includegraphics[width=1\textwidth,clip]{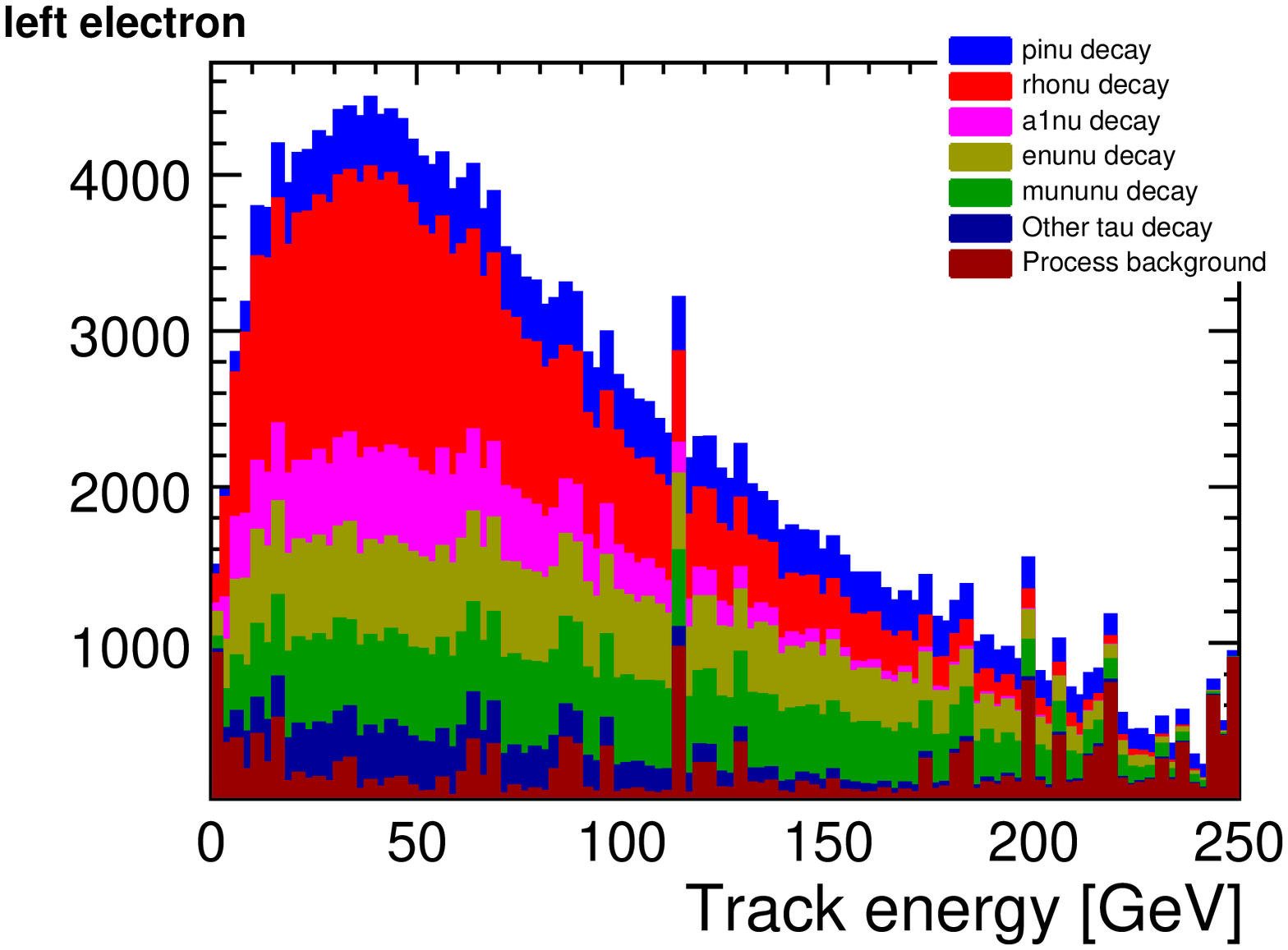}\par (c)
			\includegraphics[width=1\textwidth,clip]{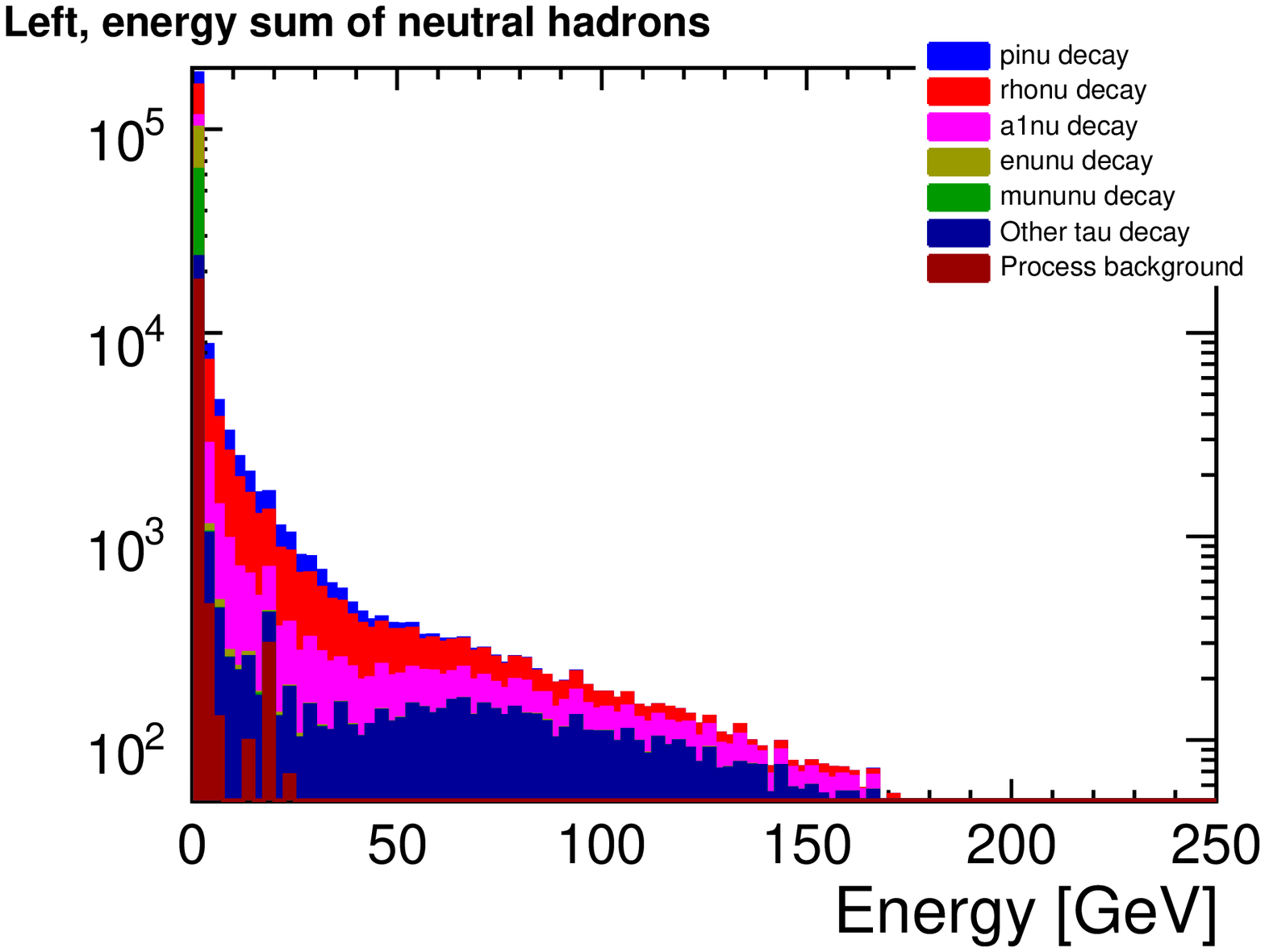}\par (e)
			\includegraphics[width=1\textwidth,clip]{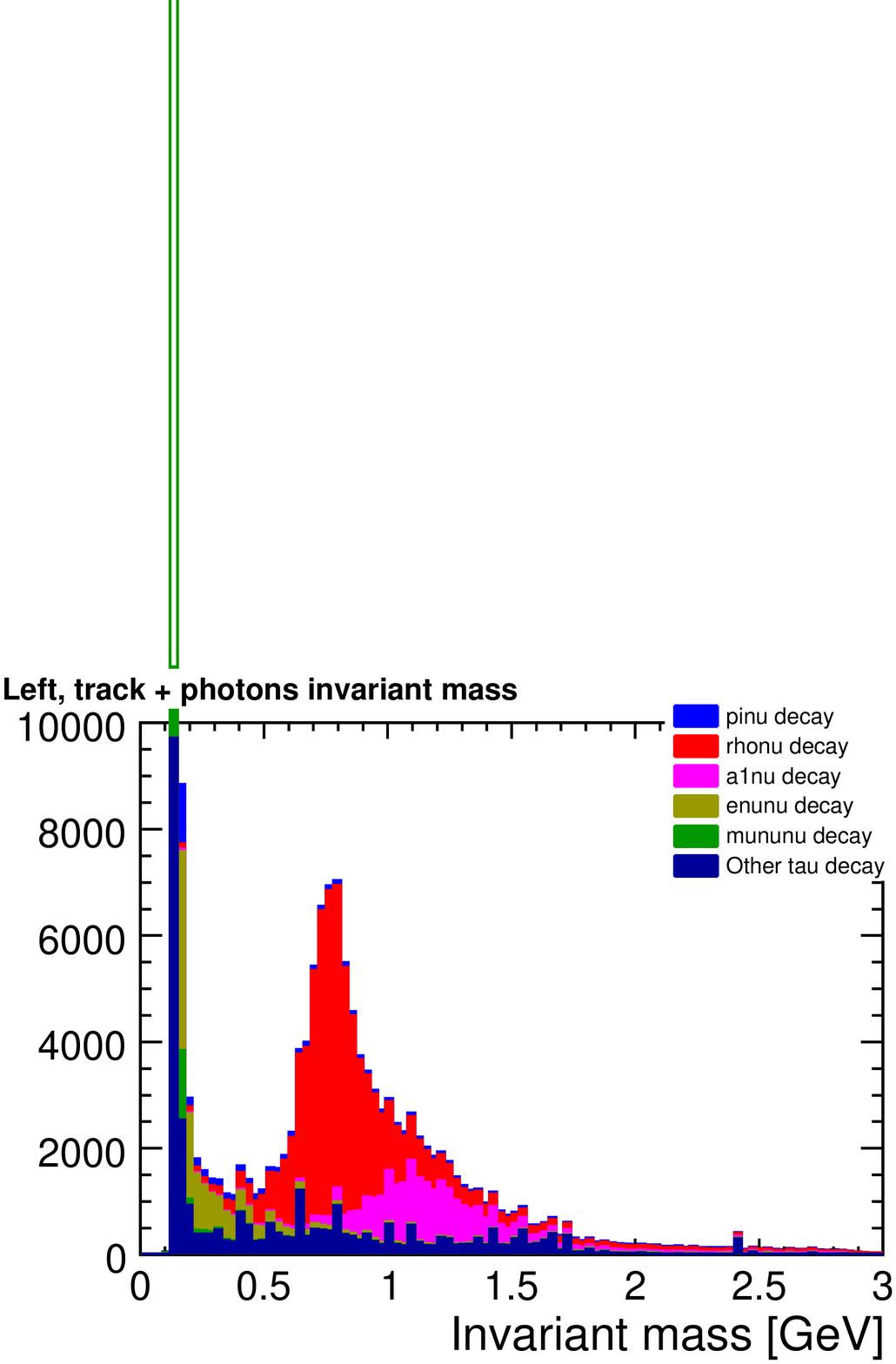}\par (g)
			\includegraphics[width=1\textwidth,clip]{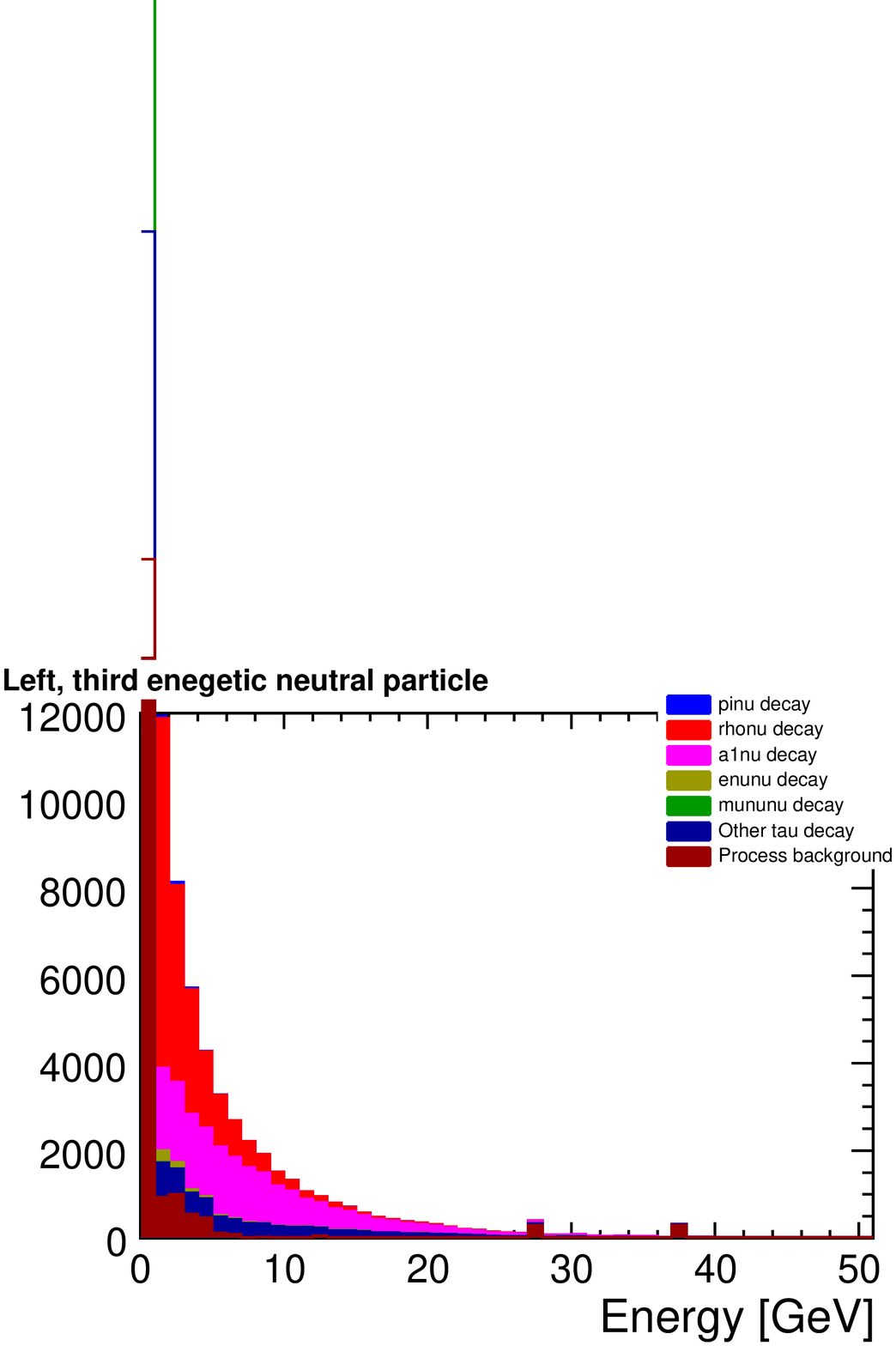}\par (i)
		\end{center}		
	\end{minipage}
	\begin{minipage}{0.35\textwidth}
		\begin{center}
			\includegraphics[width=1\textwidth,clip]{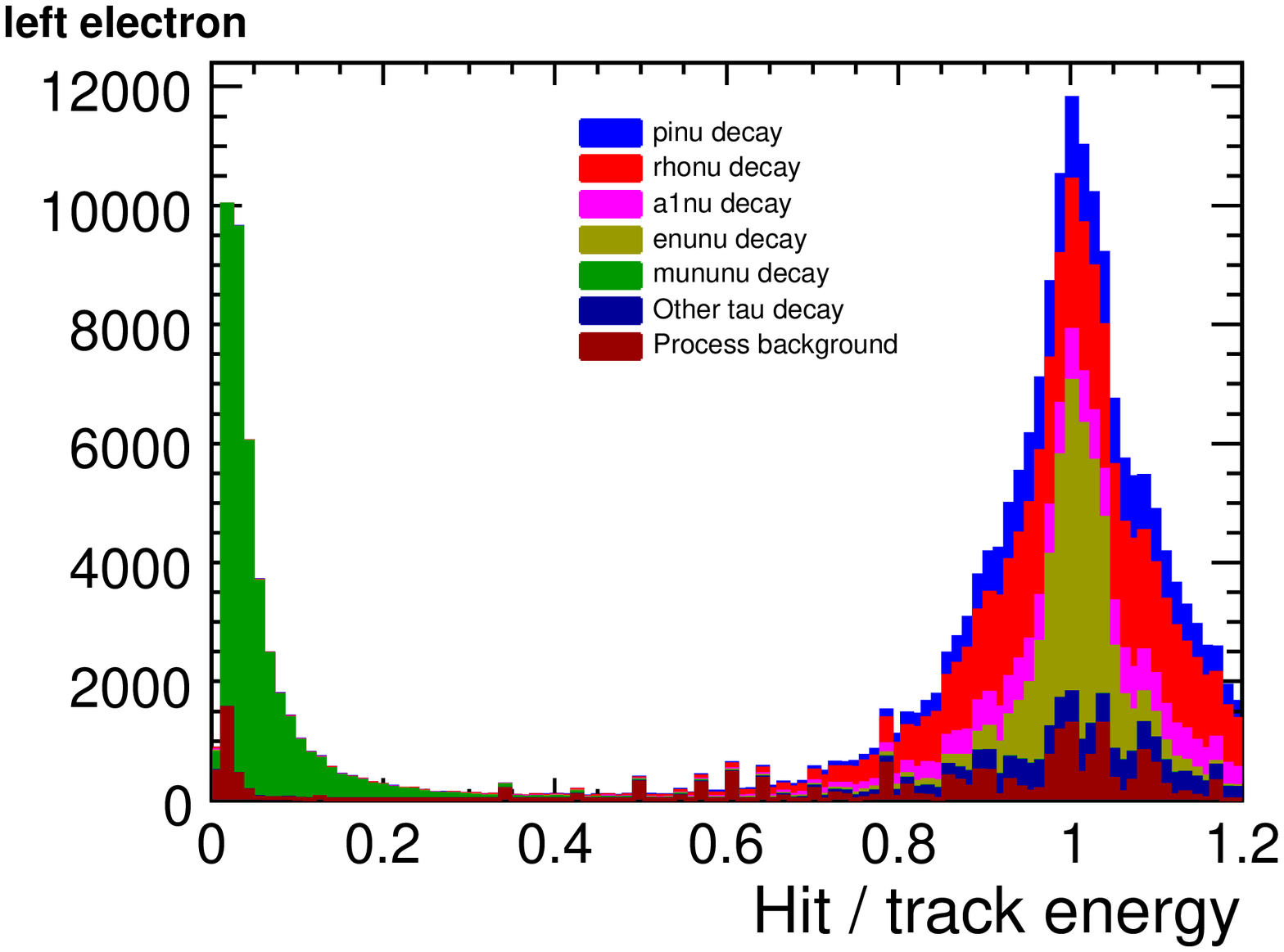}\par (b)
			\includegraphics[width=1\textwidth,clip]{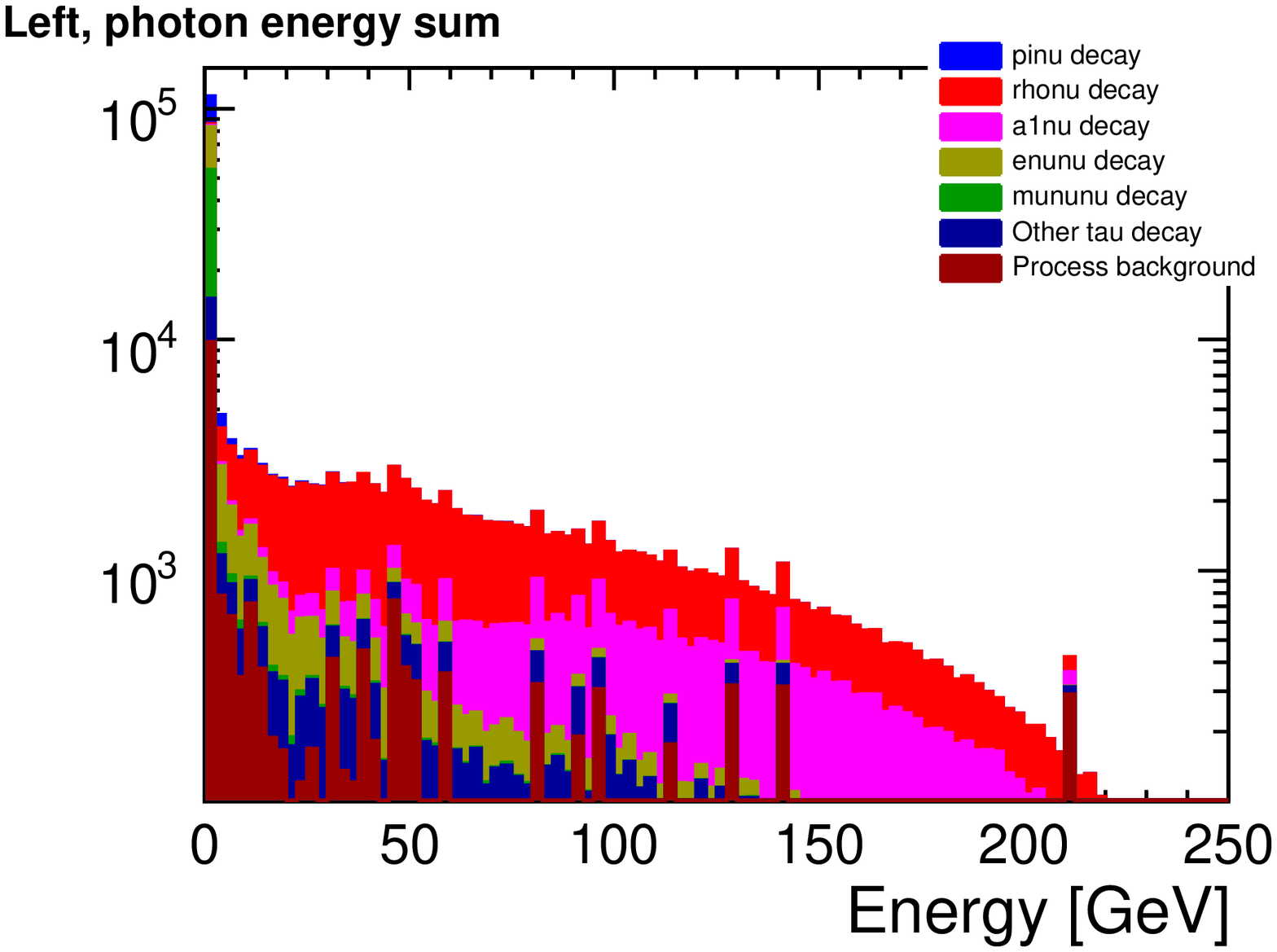}\par (d)
			\includegraphics[width=1\textwidth,clip]{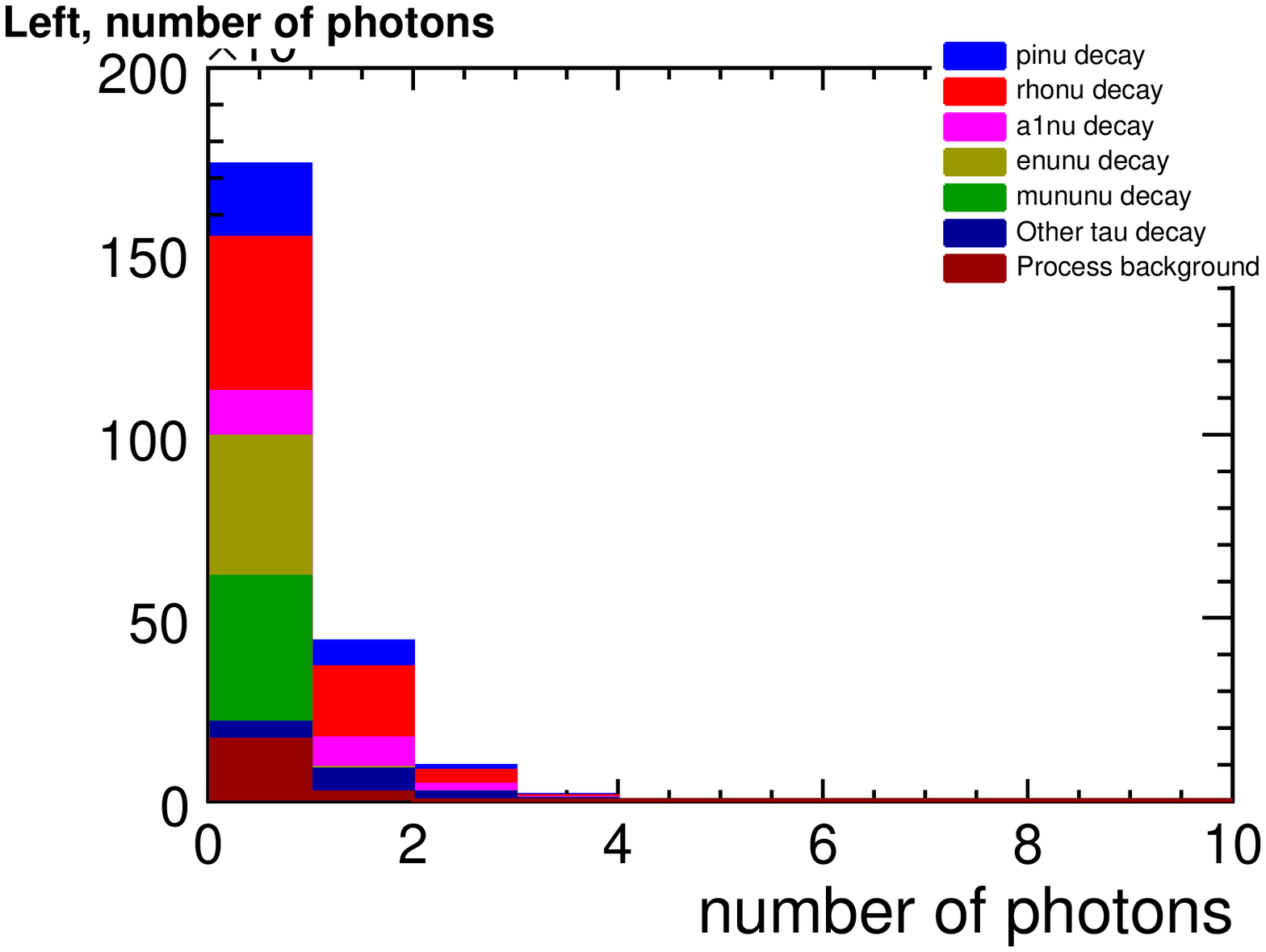}\par (f)
			\includegraphics[width=1\textwidth,clip]{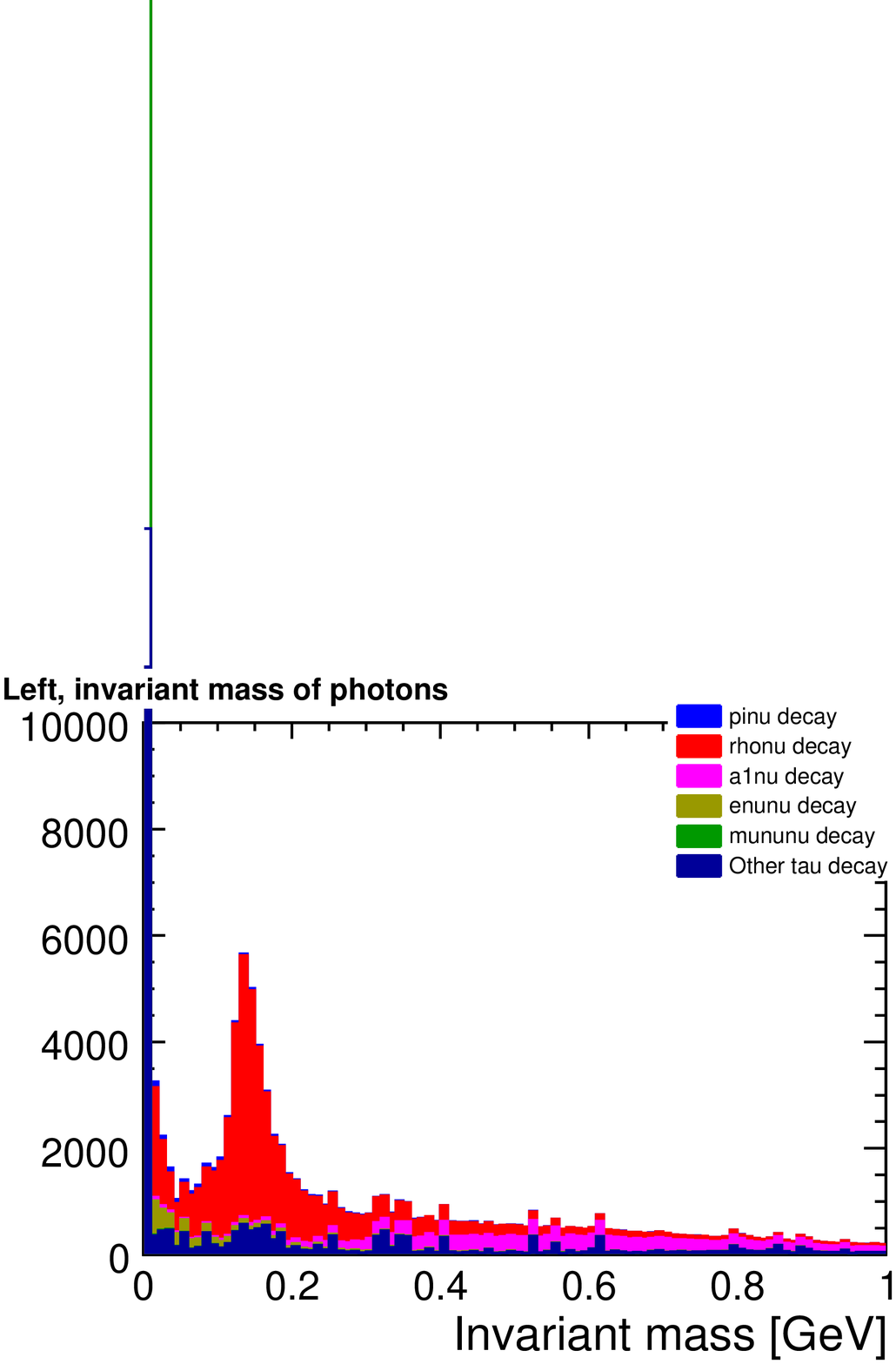}\par (h)
		\end{center}		
	\end{minipage}
	\end{center}		
	\caption{Distributions of the input variables for the 1-prong neural network.
e$^-_\mathrm{L}$ (80\%) e$^+_\mathrm{R}$ polarization is used for the plots.}
	\label{fig:inputvars1p}
\end{figure}

\begin{figure}
	\begin{center}
	\begin{minipage}{0.35\textwidth}
		\begin{center}
			\includegraphics[width=1\textwidth]{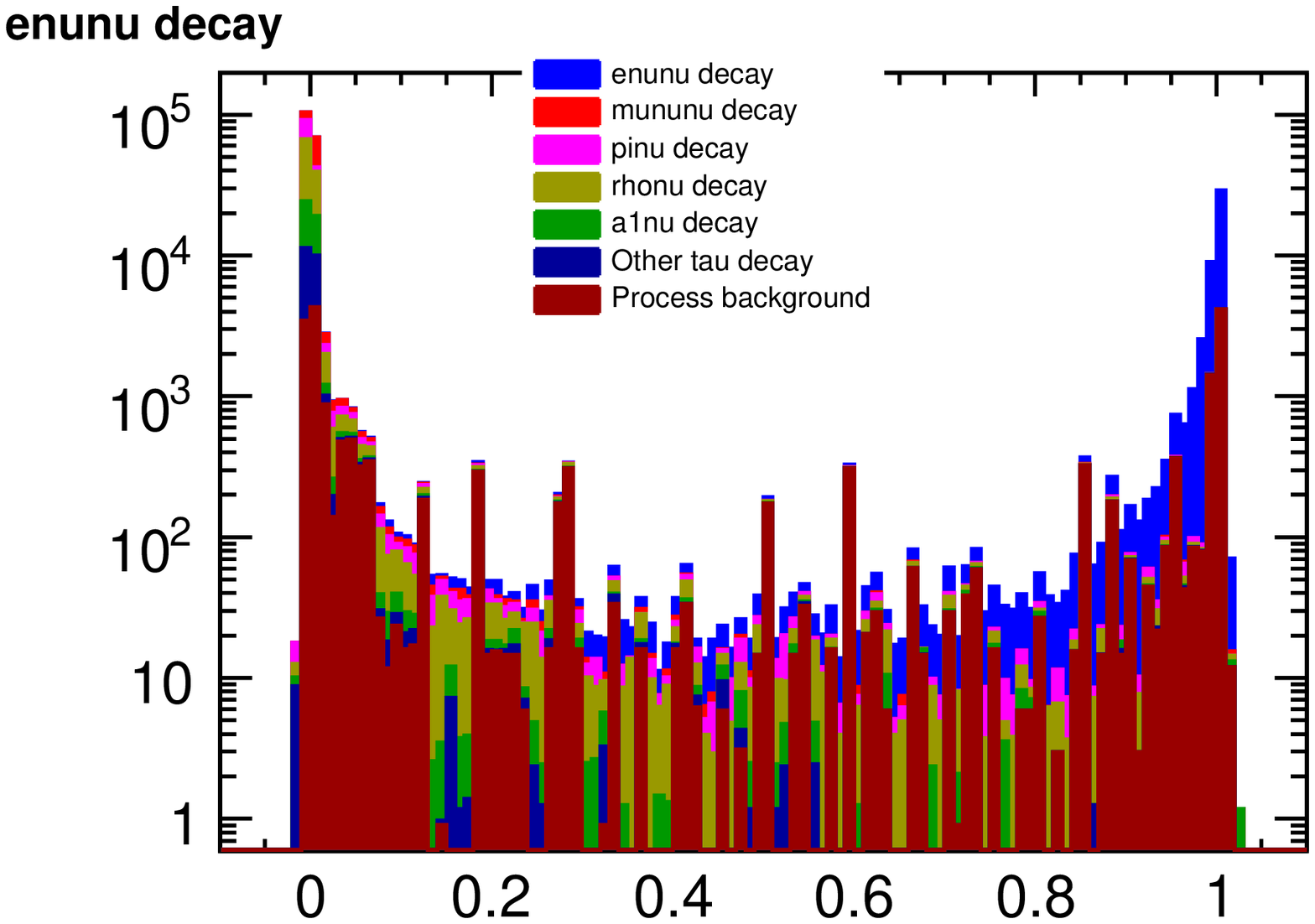}\par (a)
			\includegraphics[width=1\textwidth]{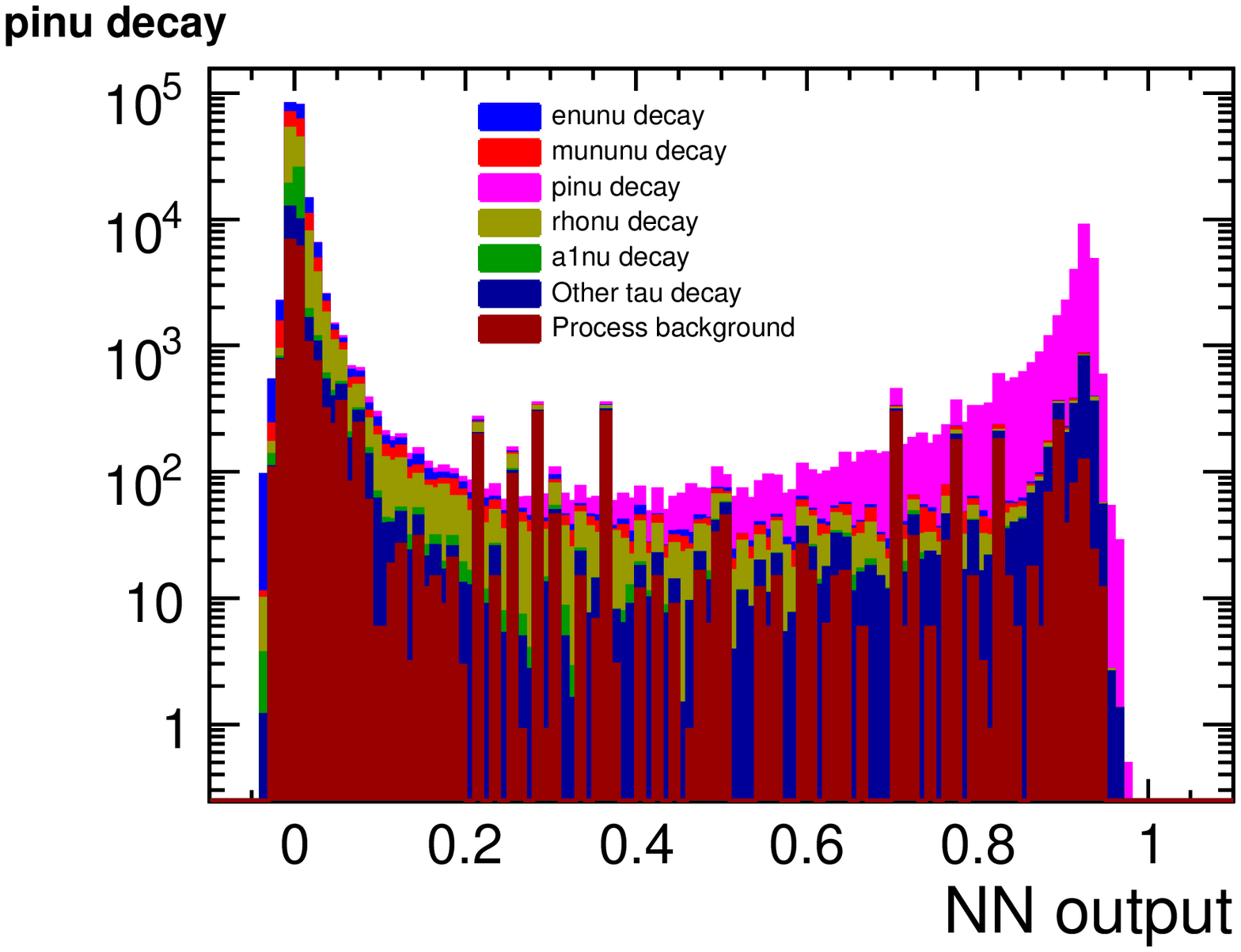}\par (c)
			\includegraphics[width=1\textwidth]{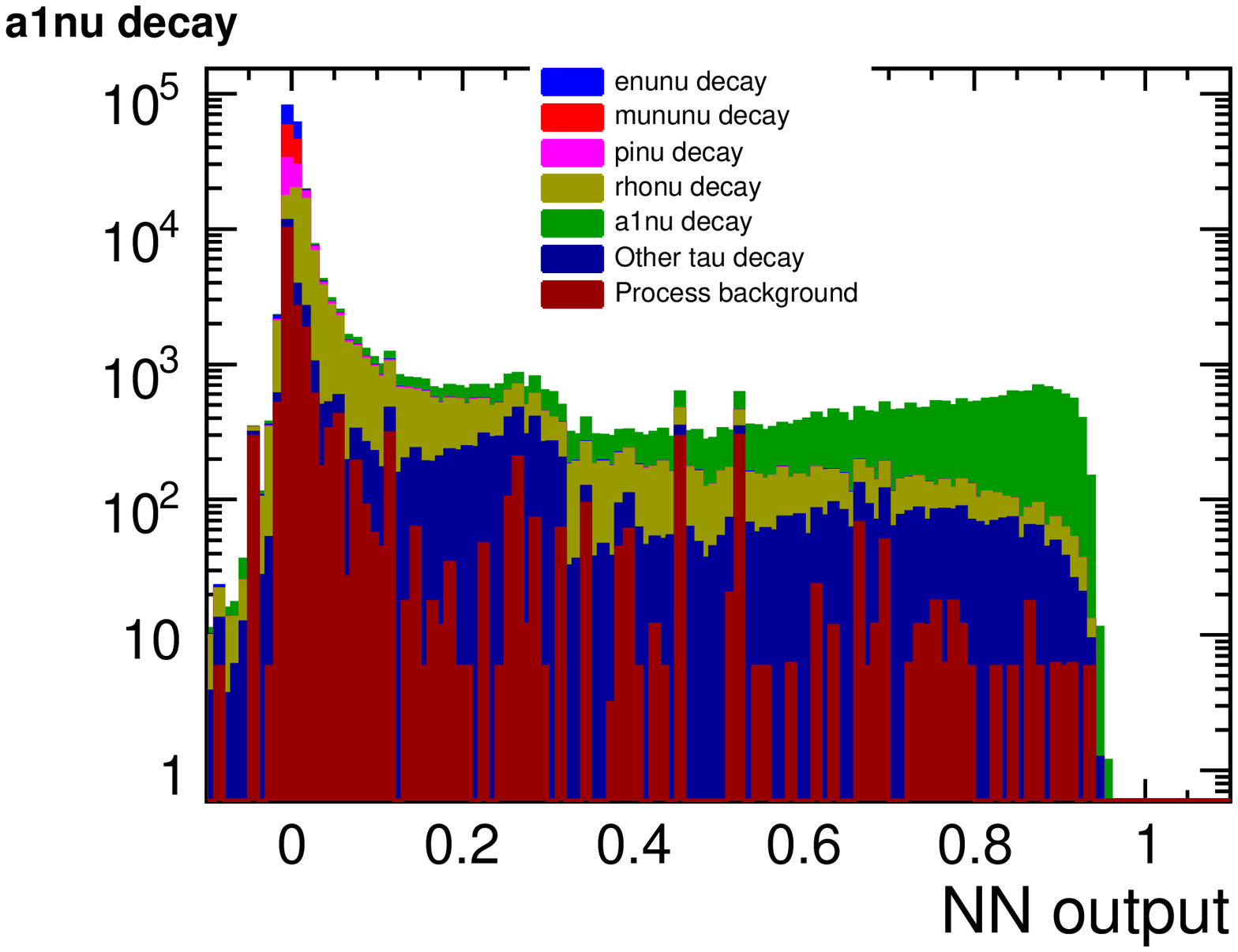}\par (e)
		\end{center}		
	\end{minipage}
	\begin{minipage}{0.35\textwidth}
		\begin{center}
			\includegraphics[width=1\textwidth]{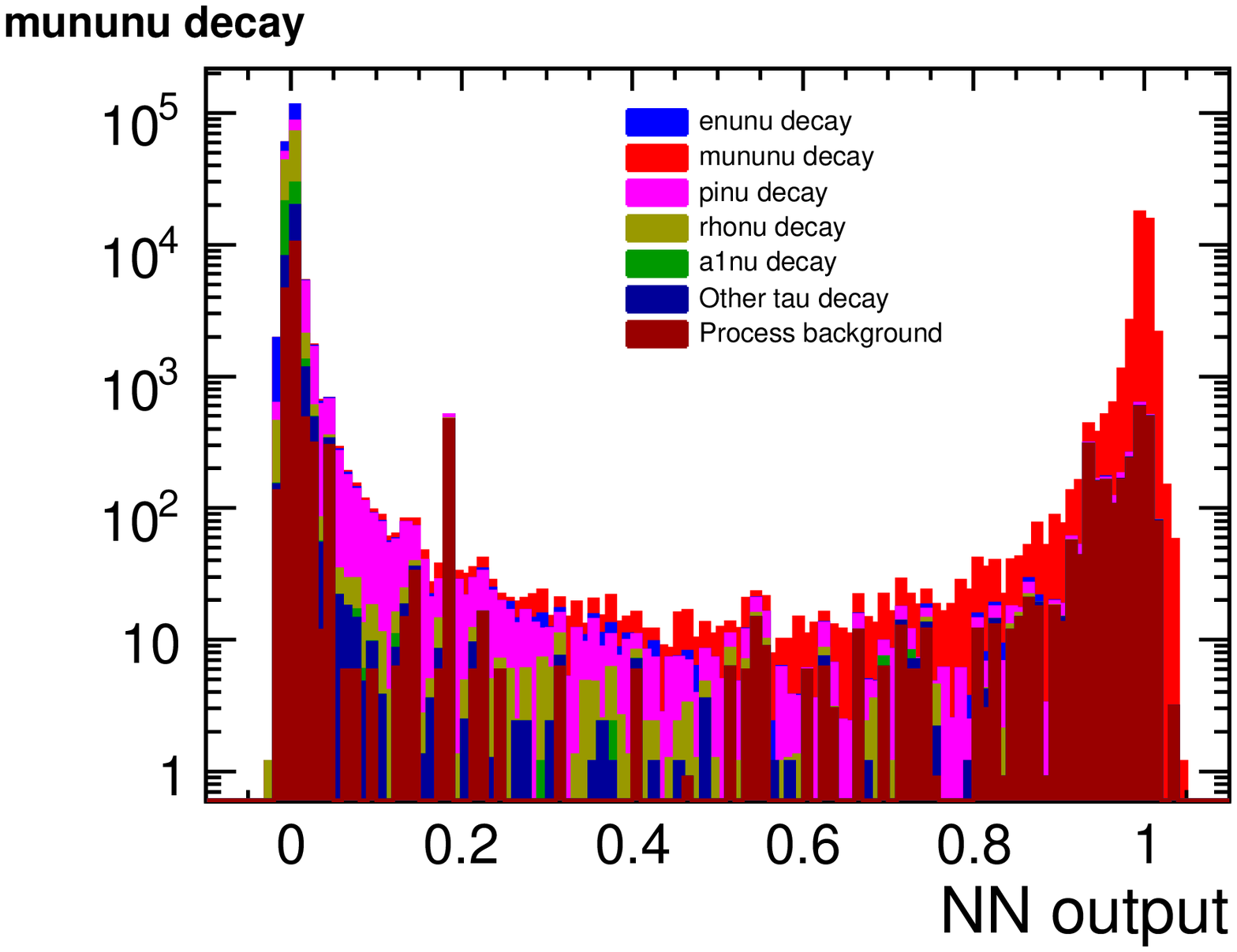}\par (b)
			\includegraphics[width=1\textwidth]{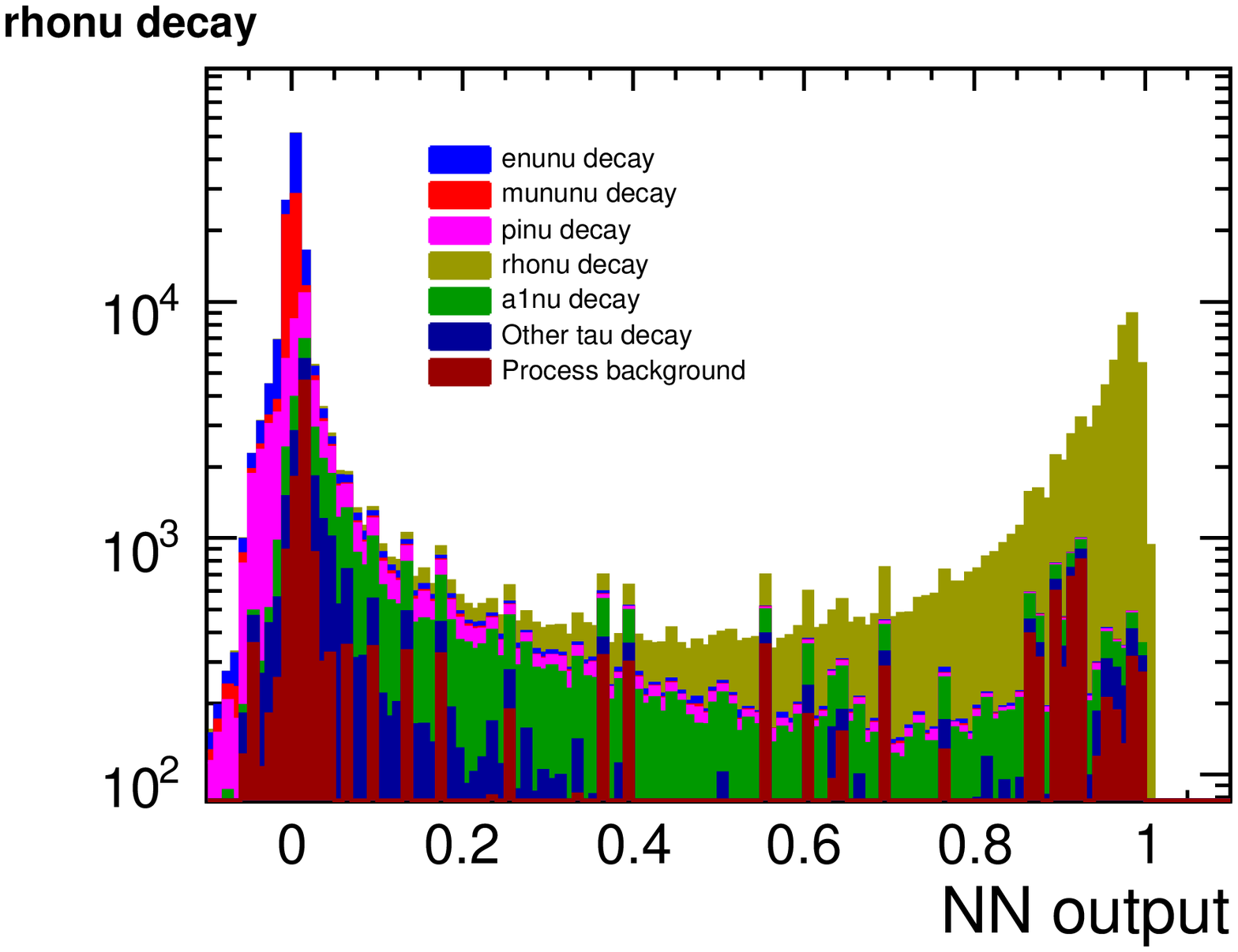}\par (d)
			\includegraphics[width=1\textwidth]{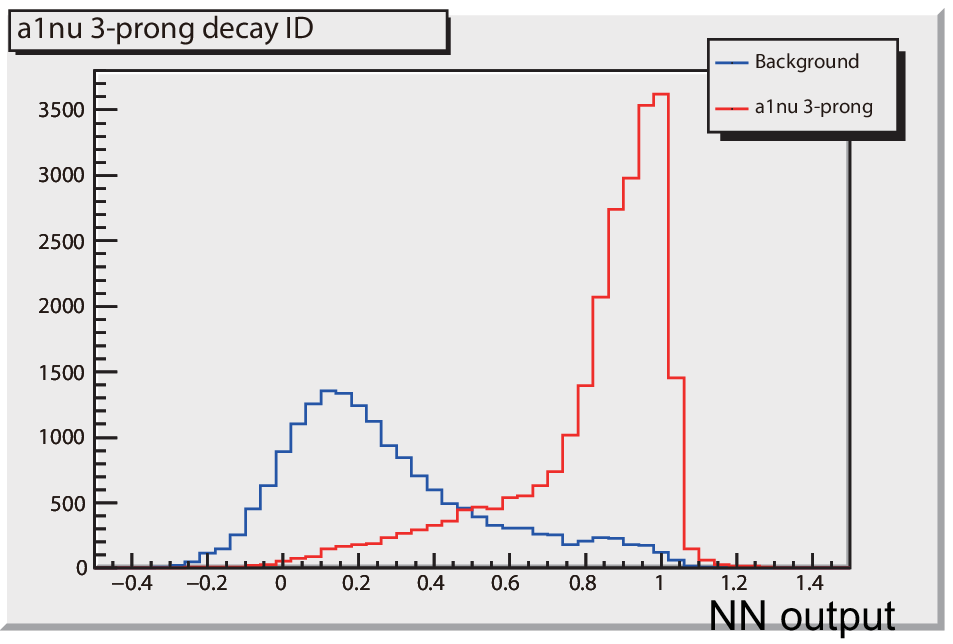}\par (f)
		\end{center}		
	\end{minipage}
	\end{center}		
	\caption{Output variables for the neural net selection.
	(a)-(e) are the output of the 1-prong neural net for
	$\mathrm{e}^+\overline{\nu_\mathrm{e}}\nu_\tau$, $\mu^+\overline{\nu_\mu}\nu_\tau$,
	$\pi^+\nu_\tau$, $\rho^+\nu_\tau$ and $\mathrm{a}_1^+\nu_\tau$ modes, respectively.
	(f) is the output of the 3-prong neural net for $\mathrm{a}_1^+\nu_\tau$ identification.
	e$^-_\mathrm{L}$ (80\%) e$^+_\mathrm{R}$ polarization is used for the plots.}
	\label{fig:nnoutput}
\end{figure}

Fig.~\ref{fig:inputvars1p} and \ref{fig:inputvars3p} shows distributions of the input variables, and
Fig.~\ref{fig:nnoutput} shows distribution of the output neurons.
The mode selection is applied based on the values of the output neurons, as follows.
\begin{itemize}
	\item If one or more of the values of the output neurons exceed 0.5, The neurons which gives the highest output value
	is used as the selection.
	\item If no output values exceed 0.5, the event is classified as `others'.
\end{itemize}

Table \ref{tbl:nnselection} shows the obtained efficiency and purity for the mode selection.
These values are obtained with the half of the tau-pair events which are not used for the training.
$\gtrsim$ 90\% efficiency and purity is obtained for all decay modes except 1-prong a$_1$ decay.

\begin{table}
\begin{center}
\begin{tabular}{|l|r|r|}
\hline
Modes																	& Purity	& Efficiency \\ \hline\hline
$\mathrm{e}\nu\nu$										& 98.9\%	& 98.9\% \\\hline
$\mu\nu\nu$														& 98.8\%	& 99.3\% \\\hline
$\pi\nu$															& 96.0\%  & 89.5\% \\\hline
$\rho\nu$															& 91.6\%  & 88.6\% \\\hline
$a_1\nu$ (1-prong)										& 67.2\%	& 73.4\% \\\hline
$a_1\nu$ (3-prong)										& 91.1\%	& 88.9\% \\\hline
\end{tabular}
\caption{Purity and efficiency of the tau decay mode selection with neural networks.
Process background is not included in the purity \& efficiency numbers.}
\label{tbl:nnselection}
\end{center}
\end{table}

\section{Polarization Measurement}

\subsection{Optimal Observable}

To identify tau polarization, optimal observables\cite{optimaltau} are used for
$\mathrm{e}^+\overline{\nu_\mathrm{e}}\nu_\tau$, $\mu^+\overline{\nu_\mu}\nu_\tau$,
	$\pi^+\nu_\tau$ and $\rho^+\nu_\tau$ decay modes.
Decay distribution of all tau decay can be described as the same form,
\begin{equation}
	W = \frac{1}{2}(1 + p\cos\theta_h)
\end{equation}
where $p$ is polarization of $\tau$ (-1 to 1) and $\theta_h$ is the opening angle of
polarimator vector $\vec{h}$ with respect to the $\tau$ momentum vector.
Explicit notation of $\vec{h}$ varies by the decay modes:
for pure-leptonic decay, flight direction of antineutrino is $\vec{h}$ and
for $\pi^+\nu_\tau$ decay, flight direction of pion is $\vec{h}$.
For the multipion decay, $\vec{h}$ is constructed from the hadronic current.

To reconstruct $p$ from a set of observables, we split $W$ to $p$-dependent
and $p$-independent components such as
\begin{equation}
	W(\vec{\xi}) = f(\vec{\xi}) + pg(\vec{\xi}),
\end{equation}
and the optimal observable $\omega$ is defined as
\begin{equation}
	\omega = \frac{g(\vec{\xi})}{f(\vec{\xi})}.
\end{equation}
By definition, probability density $P$ at $\omega$ for the polarization $p$ gives
\begin{equation}
	\frac{P(\omega;p) - P(\omega;p=0)}{P(\omega;p=0)} = \omega
\end{equation}
and $p$ can be easily obtained from the $\omega$ distribution.

The explicit formula of $\omega$ for each decay mode is as follows\cite{optimaltau2}.
\begin{enumerate}
	\item Pure-leptonic decay:
	\begin{equation}
		\omega_\ell = \frac{1 + x - 8x^2}{5 + 5x - 4x^2}
	\end{equation}
	where $x$ is the lepton energy divided by $\tau$ energy (250 GeV in this case).
	Since the pure-leptonic decay mode has two missing neutrinos,
	polarization discrimination power is weaker than semi-leptonic decay modes.
	\item $\pi^+\nu_\tau$ decay:
	\begin{equation}
		\omega_\pi = 2x - 1.
	\end{equation}
	This mode has maximum polarization discrimination power since $\vec{h}$
	can be fully reconstructed.
	\item $\rho^+\nu_\tau$ decay:
	This decay mode has multiple observable particles and thus more complicated formula
	to describe $\omega$. Tau momentum direction is unobservable in this decay, so it is integrated out
	in the $\omega$ formulation.
	The explicit formula is:
	\begin{eqnarray}
	\omega_\rho &=& {\scriptstyle\frac{\left(-1 + \frac{m_\tau^2}{Q^2} + 2\left(1 + \frac{m_\tau^2}{Q^2}\right)
		\frac{3\cos^2\psi-1}{2}\frac{3\cos^2\beta-1}{2}\right)\cos\theta
		+ 3\sqrt{\frac{m_\tau^2}{Q^2}}\frac{3\cos^2\beta-1}{2}\sin2\psi\sin\theta}
		{2 + \frac{m_\tau^2}{Q^2} - 2\left(1-\frac{m_\tau^2}{Q^2}\right)\frac{3\cos^2\psi-1}{2}\frac{3\cos^2\beta-1}{2}}}	\\
	\cos\psi &=& \frac{x(m_\tau^2 + Q^2) - 2Q^2}{(m_\tau^2 - Q^2)\sqrt{x^2 - 4Q^2/s}} \\
	x &=& 2\frac{E_h}{\sqrt{s}}
	\end{eqnarray}
	where $E_h$ is the energy sum of $\rho$ (which equals to the cluster energy), $Q^2$ is the invariant mass of the
	visible particles (should equals to $m_\rho = 0.77$ GeV but obtained from the event),
	$\sqrt{s}$ is the center-of-mass energy (500 GeV),
	$\theta$ is the angle of the $\rho$ flight direction with repect to $\tau$ direction in $\tau$-rest frame, and
	$\beta$ is the angle of the charged pion flight direction with repect to $\rho$ direction in $\rho$-rest frame.
\end{enumerate}

\subsection{Polarization measurement}

\begin{figure}
	\begin{center}
	\begin{minipage}{0.35\textwidth}
		\begin{center}
			\includegraphics[width=1\textwidth]{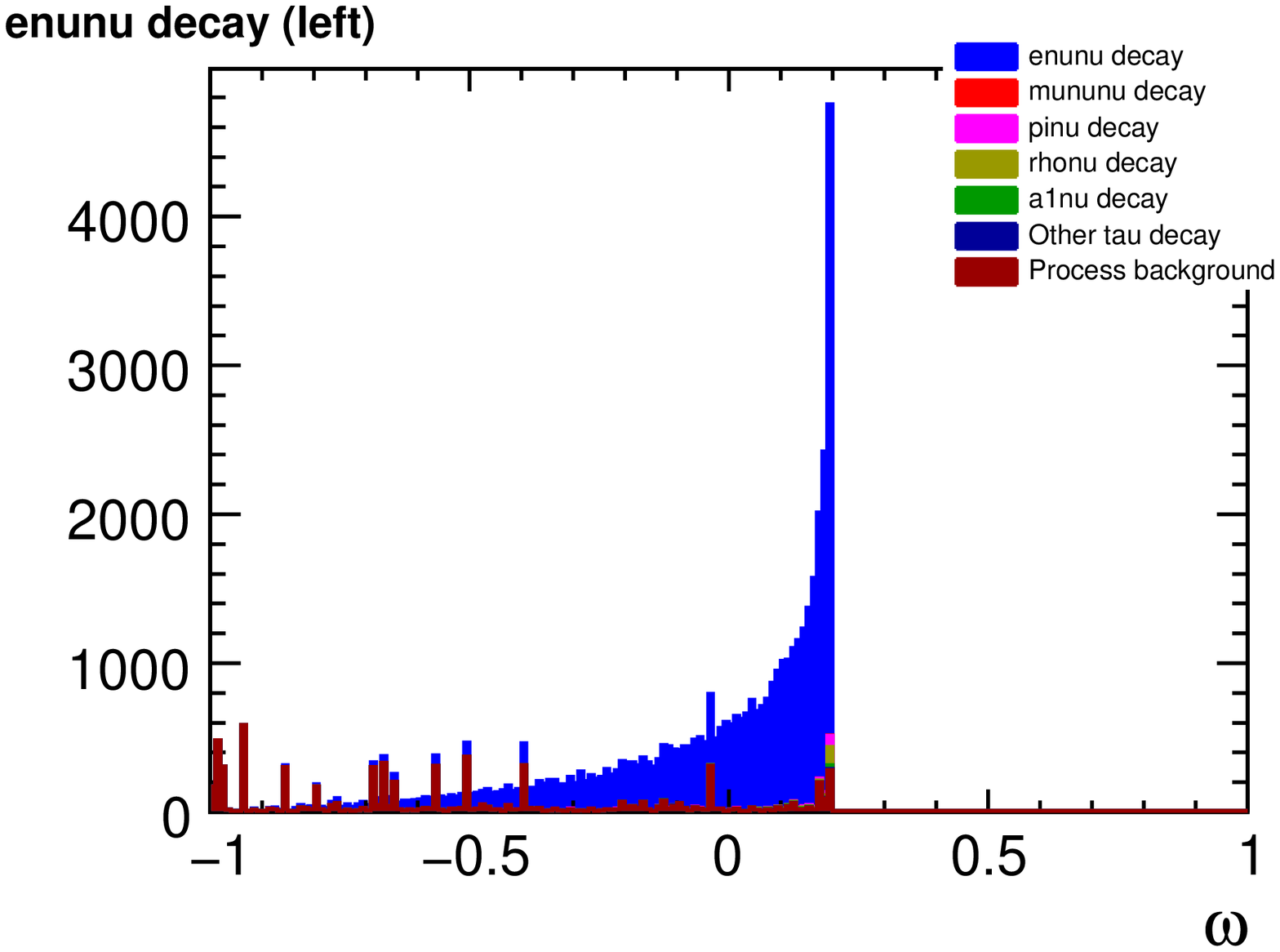}
			\includegraphics[width=1\textwidth]{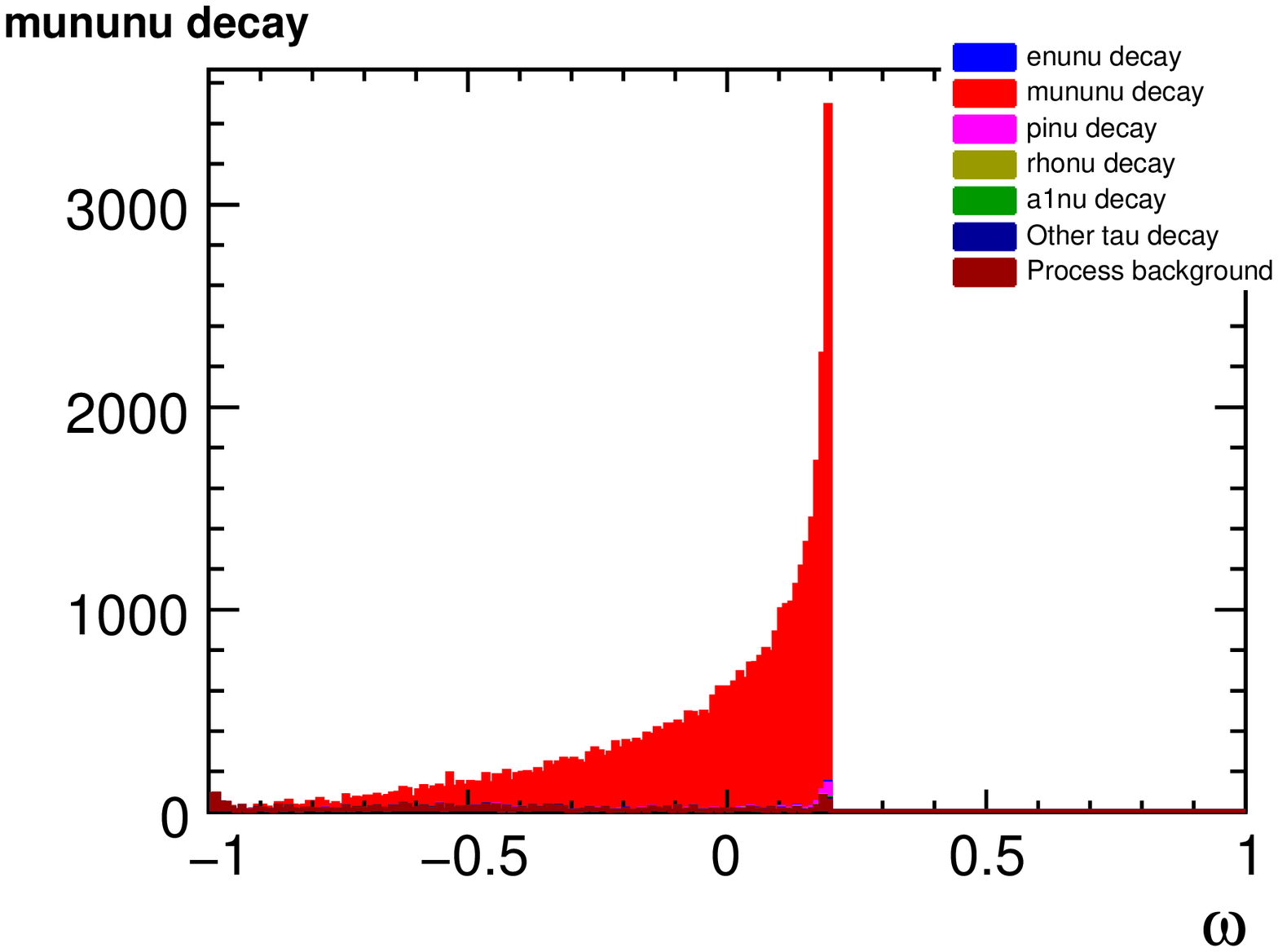}
			\includegraphics[width=1\textwidth]{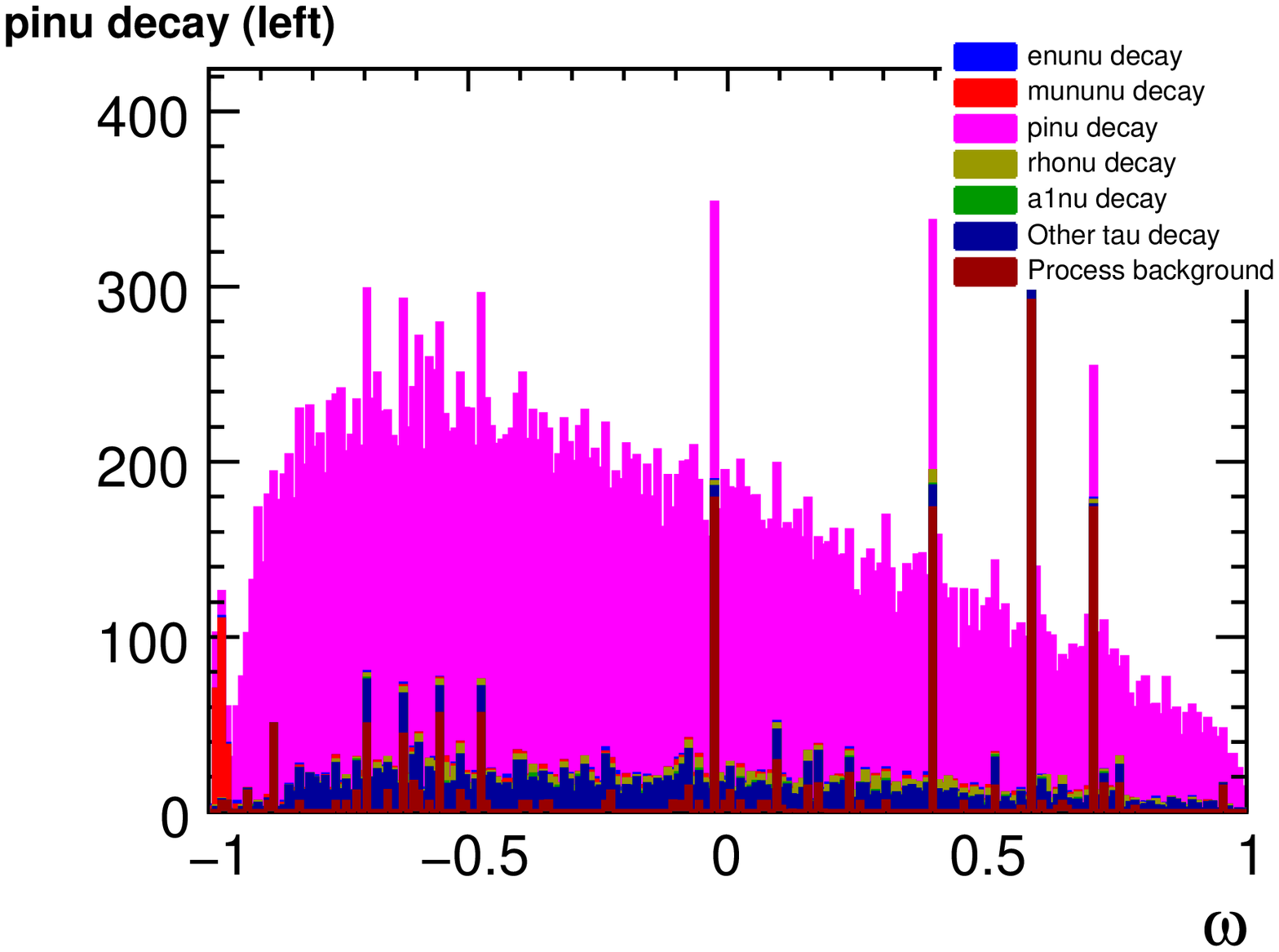}
			\includegraphics[width=1\textwidth]{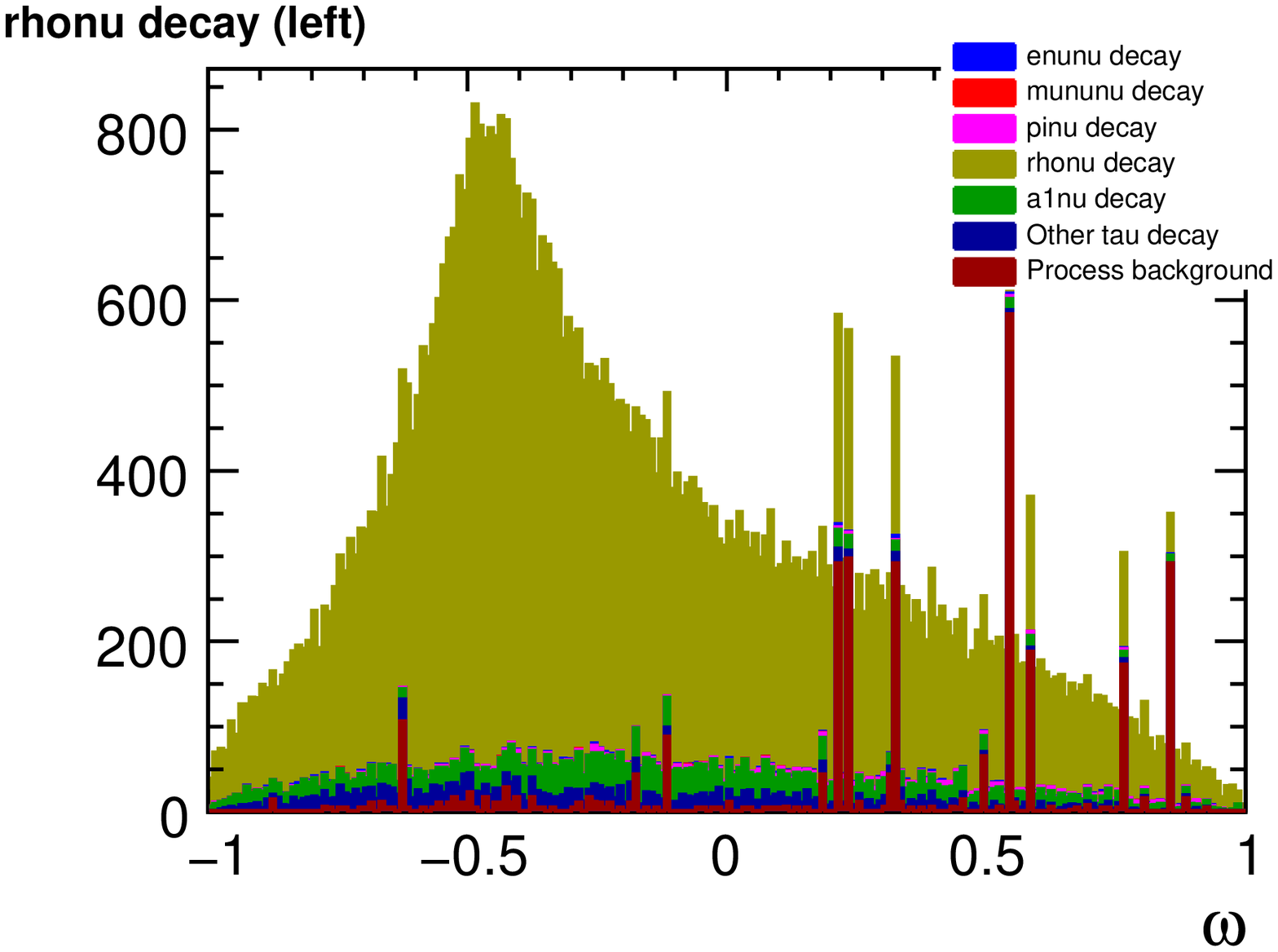}
		\end{center}		
	\end{minipage}
	\begin{minipage}{0.35\textwidth}
		\begin{center}
			\includegraphics[width=1\textwidth]{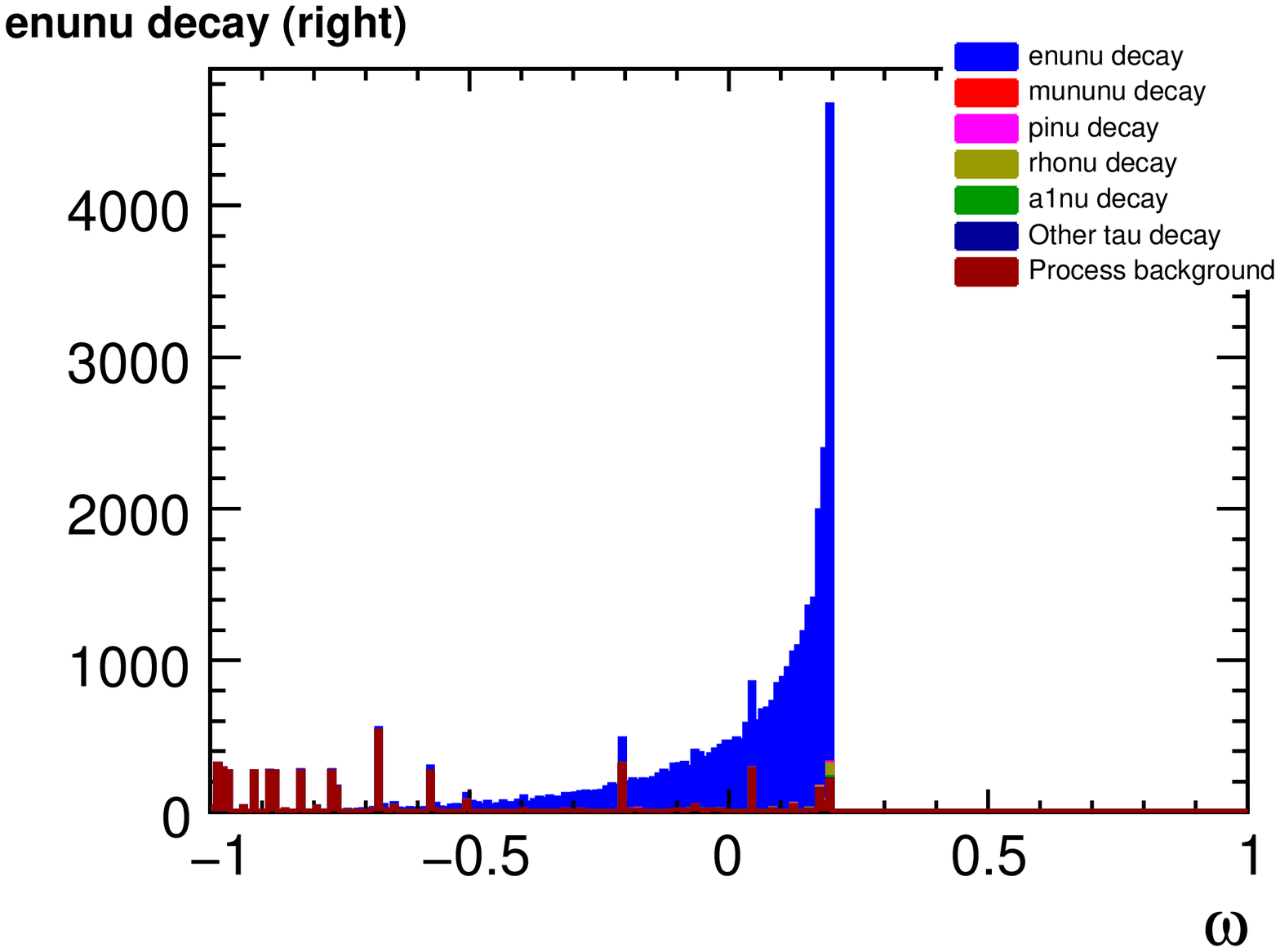}
			\includegraphics[width=1\textwidth]{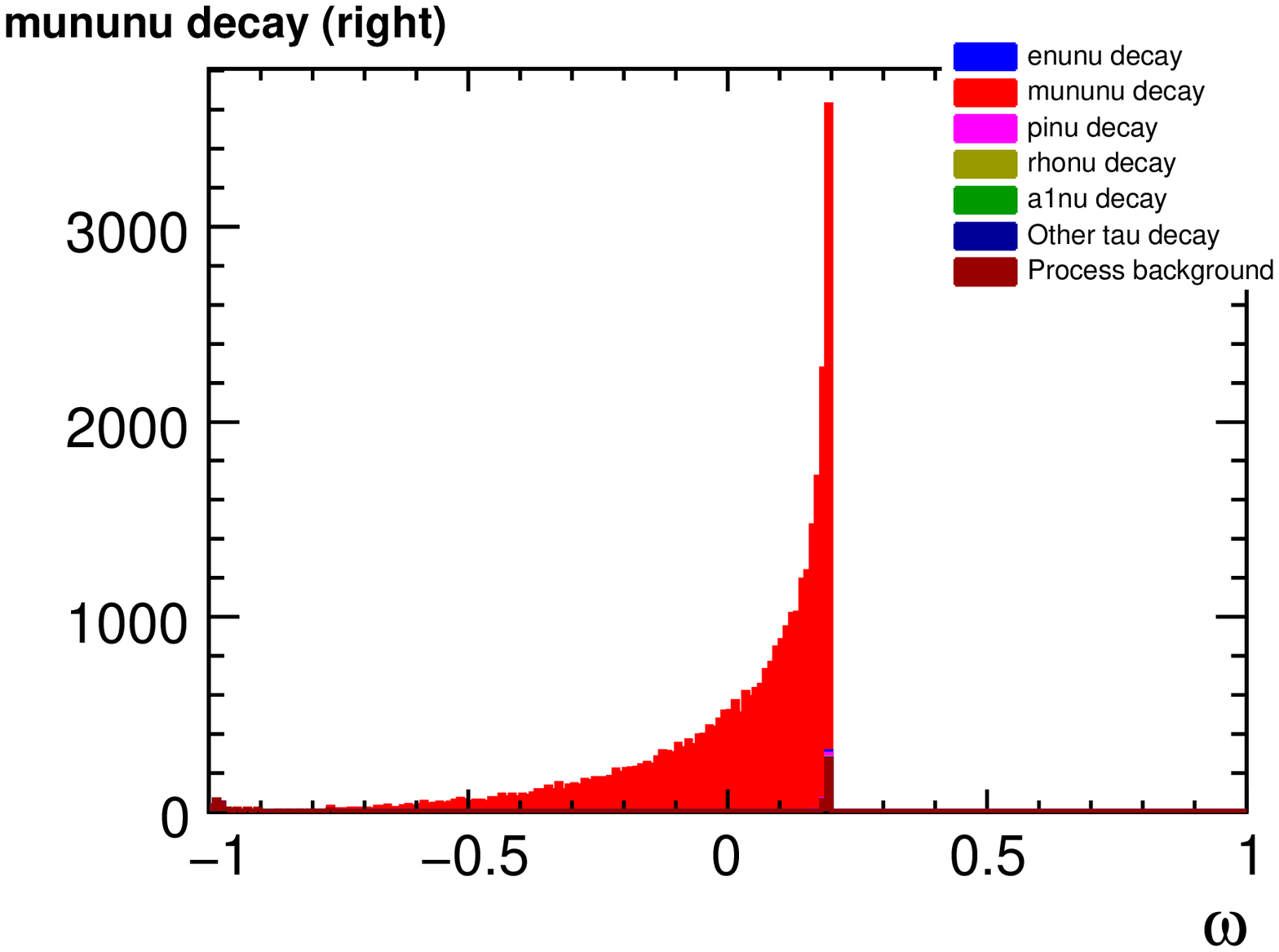}
			\includegraphics[width=1\textwidth]{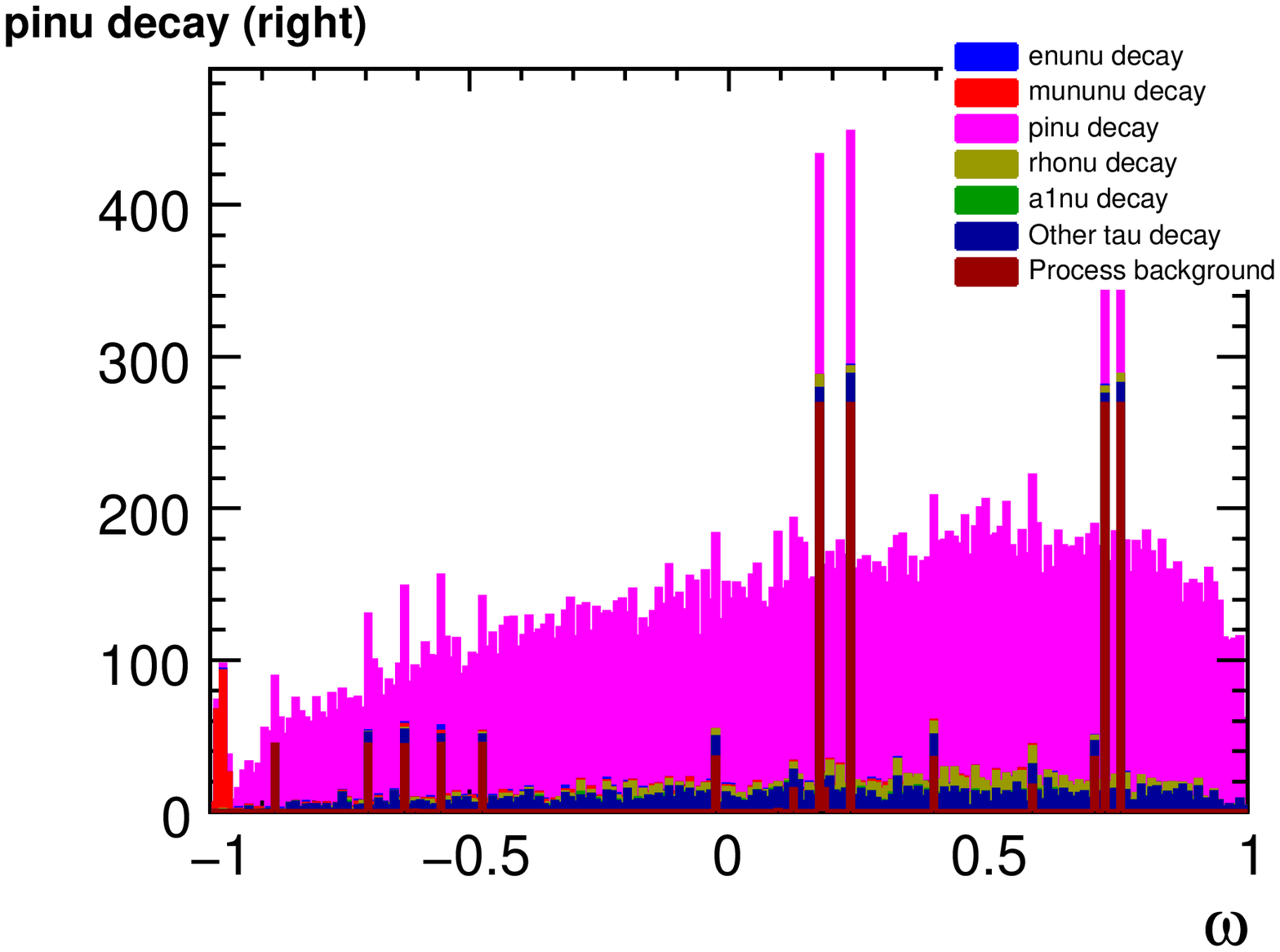}
			\includegraphics[width=1\textwidth]{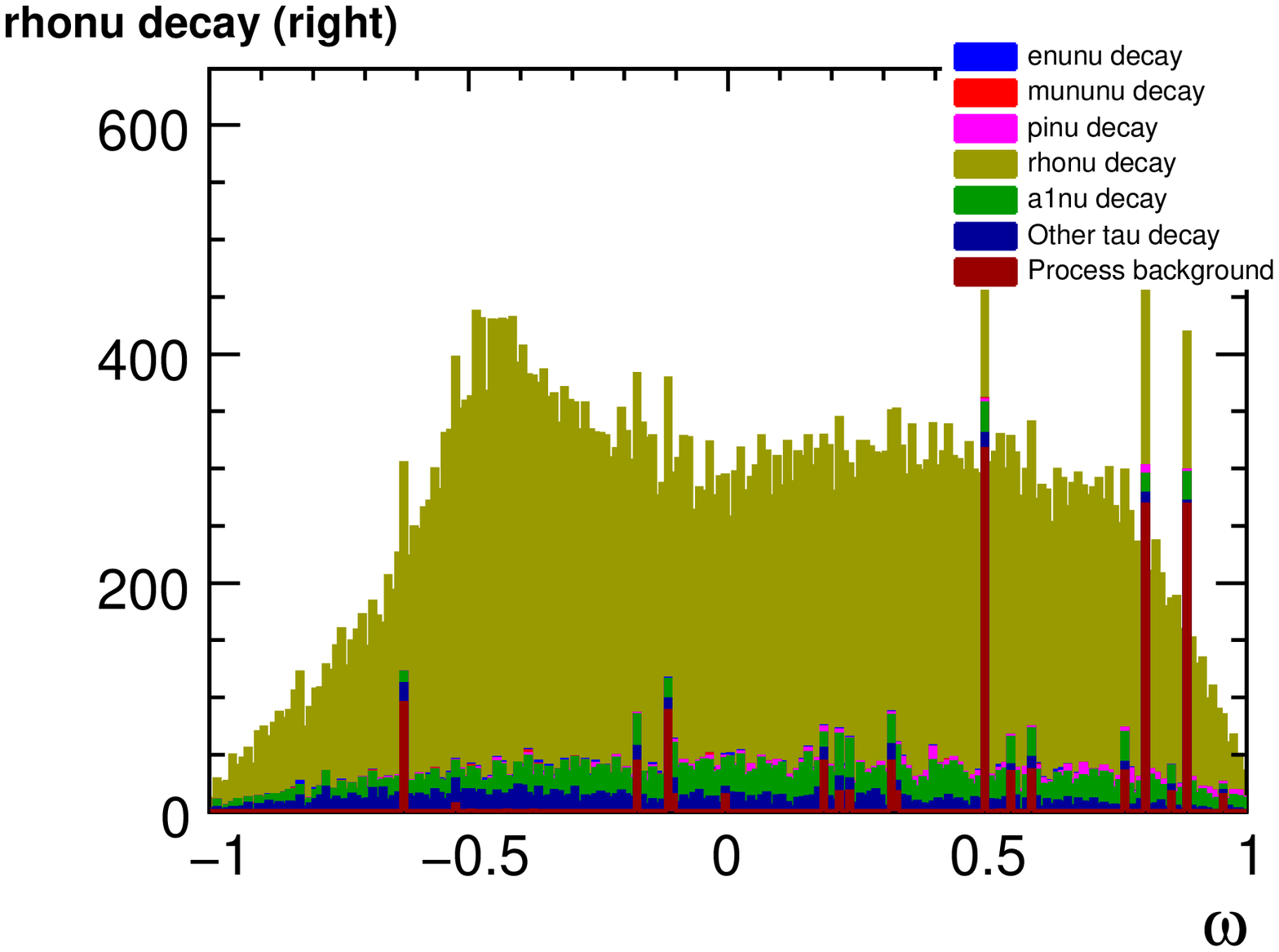}
		\end{center}		
	\end{minipage}
	\end{center}		
	\caption{Distribution of the optimal observable for each decay mode.
	The left column shows distribution of e$^-_\mathrm{L}$ (80\%) e$^+_\mathrm{R}$ (30\%) events, and
	the right column shows distribution of e$^-_\mathrm{R}$ (80\%) e$^+_\mathrm{L}$ (30\%) events.}
	\label{fig:omegadist}
\end{figure}

Figure \ref{fig:omegadist} shows the $\omega$ distribution for each decay mode passing the neural net selection.
For the leptonic mode, most of the events are concentrated on the $\omega \sim 0$ region, reflecting to the
weak discrimination power. For the $\pi^+\nu_\tau$ and $\rho^+\nu_\tau$
modes, $\omega$ distribution is broadly distributed and large difference
between left and right polarization can be seen.

Polarization $p$ can be obtained by following procedure.
\begin{enumerate}
	\item Mode and process background is eliminated from each bin of the $\omega$ histograms and
		statistical error of background remains included in the error of each bin.
	\item Histograms from all decay modes are summed into one histogram.
	\item The histogram with polarizing sample (polarization $p$) is divided by non-polarizing sample
		after normalizing both histograms.
	\item Perform linear fit passing (0,1) to the divided histogram (one parameter fit).
		Obtained slope stands for $p$.
\end{enumerate}

\begin{figure}
	\begin{center}
		\includegraphics[width=0.7\textwidth]{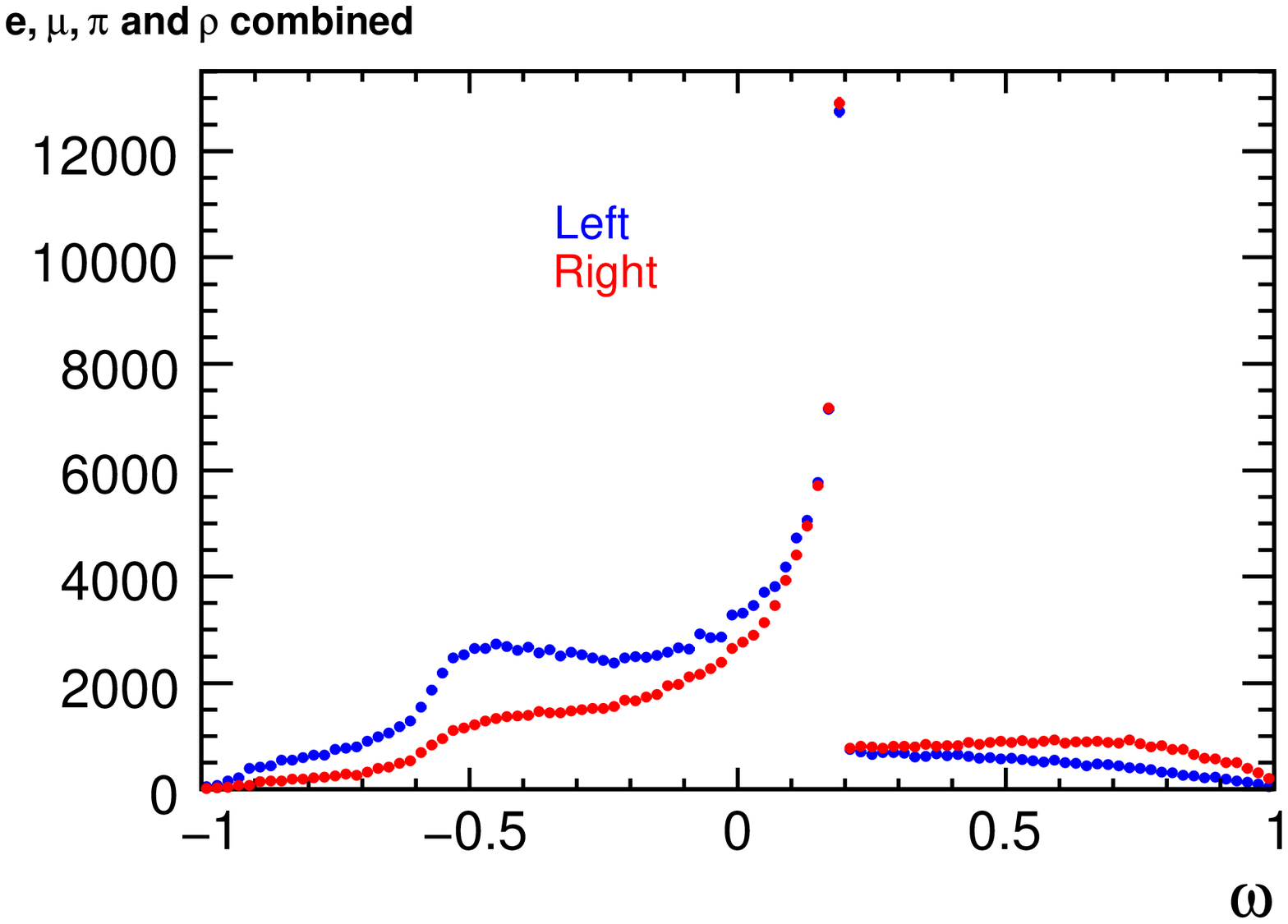}
	\end{center}		
	\caption{Distribution of the optimal observable after summing up all decay modes.}
	\label{fig:omegasum}
\end{figure}

\begin{figure}
	\begin{center}
	\begin{minipage}{0.45\textwidth}
		\begin{center}
			\includegraphics[width=1\textwidth]{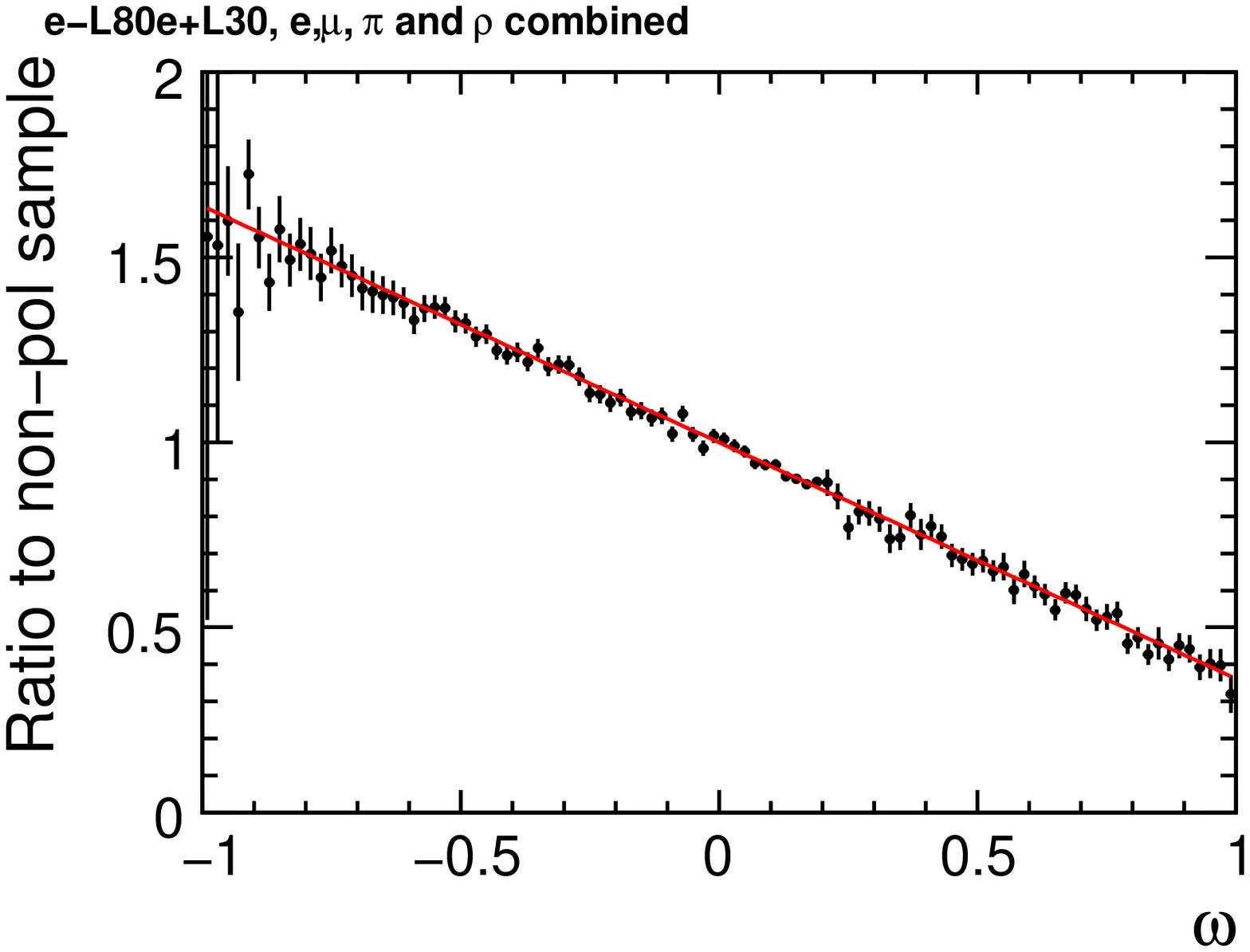}
		\end{center}		
	\end{minipage}
	\begin{minipage}{0.45\textwidth}
		\begin{center}
			\includegraphics[width=1\textwidth]{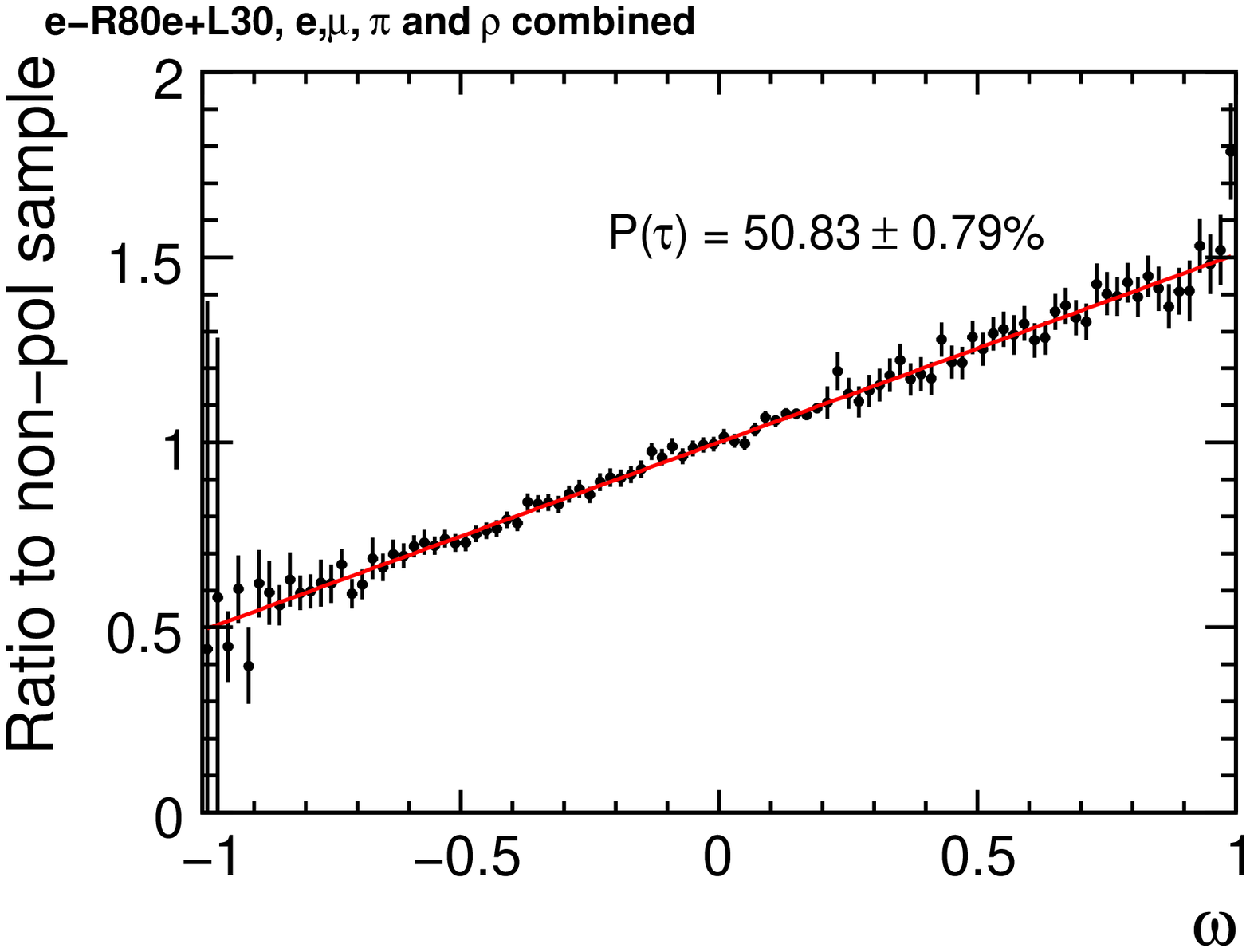}
		\end{center}		
	\end{minipage}
	\end{center}		
	\caption{Ratio of the polarizing sample to the non-polarizing sample.
		Linear fit is applied to obtain $p$ value.}
	\label{fig:omegastraight}
\end{figure}

Figure \ref{fig:omegasum} shows the combined $\omega$ distribution and
Figure \ref{fig:omegastraight} shows the linear fits to obtain $p$ value.
Obtained $p$ is $-63.82 \pm 0.66$\% (e$^-_\mathrm{L}$e$^+_\mathrm{R}$, 80\% and 30\%)
and $50.83 \pm 0.79$\% (e$^-_\mathrm{R}$e$^+_\mathrm{L}$, 80\% and 30\%).

\section{Summary}

Tau-pair process has been analysed in the ILD\_00 detector model.
After the tau selection cuts, statistical error of cross section measurement is
0.29\% (e$^-_\mathrm{L}$e$^+_\mathrm{R}$, with 80\% and 30\% polarization, respectively)
and 0.32\% (e$^-_\mathrm{R}$e$^+_\mathrm{L}$).
Process background can be suppressed to around 10\% of signal events.
Forward-backward asymmetry can be determined with 0.48\% and 0.63\% statistical error.

Polarization measurement needs separation of decay modes.
The neural net selection gives $>91$\% efficiency and $>88$\% purity
of mode selection for all major decay modes except $a_1\nu$ 1 prong mode.
Polarization analysis of $\mathrm{e}^+\overline{\nu_\mathrm{e}}\nu_\tau$,
$\mu^+\overline{\nu_\mu}\nu_\tau$, $\pi^+\nu_\tau$ and $\rho^+\nu_\tau$
decay mode is performed using the optimal observable method, and it results in
$P(\tau) = -63.82 \pm 0.66$\% (e$^-_\mathrm{L}$e$^+_\mathrm{R}$) and
$P(\tau) = 50.83 \pm 0.79$\% (e$^-_\mathrm{L}$e$^+_\mathrm{R}$).

The $\mathrm{a}_1\nu_\tau$ mode is not included in the current polarization measurement.
For the 3-prong a$_1$ decay, $\tau$ direction can be reconstructed from the vertex information
and it can improve the analysis power to the same level as $\pi^+\nu_\tau$ mode.
However, since the branching ratio of 3-prong a$_1$ decay is only about 9\%,
the expected improvement with 3-prong a$_1$ decay is about 20\%.


\begin{footnotesize}


\end{footnotesize}


\end{document}